\documentclass{aa}

\usepackage{graphicx}
\usepackage[T1]{fontenc}
\usepackage[usenames,dvipsnames]{xcolor}
\usepackage{natbib}
\usepackage[normalem]{ulem}
\usepackage{csvsimple}
\usepackage{xtab,afterpage}
\usepackage{txfonts}
\usepackage[backref=page, hyperfootnotes=true, hidelinks, colorlinks, citecolor=blue, linkcolor=blue, linktocpage, bookmarks=true]{hyperref}
\usepackage{footnote}
\usepackage{float}
\usepackage{tabularx}
\usepackage{threeparttablex}
\usepackage{enumitem}
\usepackage[bottom, flushmargin]{footmisc}

\newcolumntype{Y}{>{\centering\arraybackslash}X}

\makeatletter
\newbox\LT@firstfoot
\def\endfirstfoot{\LT@end@hd@ft\LT@firstfoot}
\newdimen\LT@footdiff
\def\LT@start{%
  \let\LT@start\endgraf
  \endgraf\penalty\z@
  \vskip\LTpre\endgraf
  \LT@footdiff-\ht\LT@foot
  \advance\LT@footdiff\ht\LT@firstfoot
  \dimen@\pagetotal
  \advance\dimen@ \ht\ifvoid\LT@firsthead\LT@head\else\LT@firsthead\fi
  \advance\dimen@ \dp\ifvoid\LT@firsthead\LT@head\else\LT@firsthead\fi
  \advance\dimen@ \ht\ifvoid\LT@firstfoot\LT@foot\else\LT@firstfoot\fi
  \dimen@ii\vfuzz
  \vfuzz\maxdimen
  \setbox\tw@\copy\z@
  \setbox\tw@\vsplit\tw@ to \ht\@arstrutbox
  \setbox\tw@\vbox{\unvbox\tw@}%
  \vfuzz\dimen@ii
  \advance\dimen@ \ht
      \ifdim\ht\@arstrutbox>\ht\tw@\@arstrutbox\else\tw@\fi
  \advance\dimen@\dp
      \ifdim\dp\@arstrutbox>\dp\tw@\@arstrutbox\else\tw@\fi
  \advance\dimen@ -\pagegoal
  \ifdim \dimen@>\z@\vfil\break\fi
  \global\@colroom\@colht
  \ifvoid\LT@firstfoot
    \ifvoid\LT@foot
    \else
      \advance\vsize-\ht\LT@foot
      \global\advance\@colroom-\ht\LT@foot
      \dimen@\pagegoal\advance\dimen@-\ht\LT@foot\pagegoal\dimen@
      \maxdepth\z@
    \fi
  \else
    \advance\vsize-\ht\LT@firstfoot
    \global\advance\@colroom-\ht\LT@firstfoot
    \dimen@\pagegoal\advance\dimen@-\ht\LT@firstfoot\pagegoal\dimen@
    \maxdepth\z@
  \fi
  \ifvoid\LT@firsthead\copy\LT@head\else\box\LT@firsthead\fi\nobreak
  \output{\LT@output}%
}
\def\LT@output{%
  \ifnum\outputpenalty <-\@Mi
    \ifnum\outputpenalty > -\LT@end@pen
      \LT@err{floats and marginpars not allowed in a longtable}\@ehc
    \else
      \setbox\z@\vbox{\unvbox\@cclv}%
      \ifdim \ht\LT@lastfoot>\ht\LT@foot
        \dimen@\pagegoal
        \advance\dimen@-\ht\LT@lastfoot
        \ifdim\dimen@<\ht\z@
          \setbox\@cclv\vbox{\unvbox\z@\copy\LT@foot\vss}%
          \@makecol
          \@outputpage
          \setbox\z@\vbox{\box\LT@head}%
        \fi
      \fi  
      \global\@colroom\@colht
      \global\vsize\@colht   
      \vbox
        {\unvbox\z@\box\ifvoid\LT@lastfoot\LT@foot\else\LT@lastfoot\fi}%
    \fi
  \else
    \ifvoid\LT@firstfoot
      \setbox\@cclv\vbox{\unvbox\@cclv\copy\LT@foot\vss}%
      \@makecol
      \@outputpage
      \global\vsize\@colroom
    \else
      \setbox\@cclv\vbox{\unvbox\@cclv\box\LT@firstfoot\vss}%
      \@makecol
      \@outputpage
      \global\advance\@colroom\LT@footdiff
      \global\vsize\@colroom
    \fi
    \copy\LT@head\nobreak
  \fi
}
\makeatother

\begin{document}

   \title{A comparison of the dynamical and model-derived parameters of the pulsating eclipsing binary KIC9850387}
   \titlerunning{Evolutionary and asteroseismic modelling of the pulsating eclipsing binary KIC9850387}


\author{
    S.~Sekaran\inst{\ref{ivs}}
    \and
    A.~Tkachenko\inst{\ref{ivs}}
    \and
    C.~Johnston\inst{\ref{ivs},\ref{conny}}
    \and
    C.~Aerts\inst{\ref{ivs}, \ref{conny}, \ref{MPI}}
}


\institute{
    Instituut voor Sterrenkunde (IvS), KU Leuven, Celestijnenlaan 200D, B-3001 Leuven, Belgium \\
    \email{sanjay.sekaran@kuleuven.be}
    \label{ivs}
    \and
    Department of Astrophysics, IMAPP, Radboud University Nijmegen, NL-6500 GL, Nijmegen, the Netherlands
    \label{conny}
    \and
    Max Planck Institute for Astronomy, Koenigstuhl 17, 69117 Heidelberg, Germany
    \label{MPI}
    }

   \date{Received December 17, 2020; accepted February 18, 2021}

 
  \abstract
   {One-dimensional (1D) stellar evolutionary models incorporate interior mixing profiles as a simplification of multi-dimensional physical processes that have a significant impact on the evolution and lifetime of stars. As such, the proper calibration of interior mixing profiles is required for the reconciliation of observational parameters and theoretical predictions. The modelling and analysis of pulsating stars in eclipsing binary systems that display gravity-mode (g-mode) oscillations allows for the precise constraints on the interior mixing profiles through the combination of spectroscopic, binary and asteroseismic obervables.}
   {We aim to unravel the interior mixing profile of the pulsating eclipsing binary KIC9850387 by comparing its dynamical parameters and the parameters derived through a combination of evolutionary and asteroseismic modelling.}
   {We created a grid of stellar evolutionary models using the stellar evolutionary code \textsc{mesa} and performed an isochrone-cloud (isocloud) based evolutionary modelling of the system. We then generated a grid of pulsational models using the stellar pulsation code \textsc{gyre} based on the age constraints from the evolutionary modelling. Finally, we performed asteroseismic modelling of the observed $\ell=1$ and $\ell=2$ period-spacing patterns, utilising different combinations of observational constraints, merit functions, and asteroseismic observables to obtain strong constraints on the interior properties of the primary star.}
   {Through a combination of asteroseismic modelling and dynamical constraints, we found that the system comprises two main-sequence components at an age of $1.2\pm0.1$\,Gyr. We also observed that asteroseismic modelling provided stronger constraints on the interior properties than evolutionary modelling. Overall, we found high levels of interior mixing, when compared to similar studies, for the primary star. We posited that this is a result of intrinsic non-tidal mixing mechanisms due to a similar observed behaviour in single stars. We investigated the high-frequency regime of KIC9850387 and found evidence of the surface effect due to the systematic frequency offset of the theoretical modes from the nearest observed modes. We also found evidence of rotational splitting in the form of a prograde-retrograde dipole g$_{1}$ mode doublet with a missing zonal mode, implying an envelope rotational frequency that is three times higher than the core rotational frequency and about 20 times slower than the orbital frequency, but we note that this result is based completely on the rotational splitting of a single dipole mode.}
   {We find that the dynamical parameters and the parameters extracted from the asteroseismic modelling of period-spacing patterns are only barely compliant, reinforcing the need for homogeneous analyses of samples of pulsating eclipsing binaries that aim to calibrating interior mixing profiles.}


   \maketitle

\section{Introduction}
\label{sec: intro}

Stellar evolutionary models are the realisation of our understanding of stellar structure and evolution, and they are the foundation for more complex models and simulations of stellar populations, clusters, and galaxies \citep{SC2005}. As such, any inaccuracies in the mathematical description or calibration of the various input physics that are included in these models inevitably propagate to the results of studies that rely on them as input. In addition, the one-dimensional (1D) nature of stellar evolutionary models necessitates the inclusion of parametric simplifications of multi-dimensional physical processes, such as the 1D approximation of turbulent convection through the incorporation of the mixing length theory \citep{BV1958}.

Stellar evolutionary models are therefore limited by the framework of these simplifications, complicating the task of improving these models. However, it has become increasingly clear in recent decades that one of the dominant sources of uncertainty regulating the evolutionary pathways of stars is the amount and prescriptions of the various mixing processes occurring in stellar interiors due to their ability to modify the amount of fuel available for nuclear fusion \citep{Maeder2009,Brott2011,Langer2012,Meynet2013}. The influence of these mechanisms is particularly significant for the evolution of stars that are born with or that develop a convective core on the main sequence ($M\gtrsim 1.15 \ \text{M}_{\odot}$): Uncertainties in the calibration and implementation of these mixing processes propagate to later evolutionary stages, most ubiquitously to the red-giant branch for intermediate-mass stars with $1.15 \ \text{M}_{\odot} \lesssim M\lesssim 2.5 \ \text{M}_{\odot}$ \citep{Constantino2015,SA2020} and to the pre-supernova stage for the massive stars \citep{Martins2015,Bowman2020}.

Internal chemical mixing is a result of independent or coupled physical mechanisms operating in different regions of the stellar interior (see \citealt{SC2017} for a review) and can be broadly divided into two classes for the $M\gtrsim 1.15 \ \text{M}_{\odot}$ regime: core-boundary mixing (CBM) and envelope mixing. CBM is an collective term that includes all prescriptions for the transport of material from the convective core into the radiative envelope (or vice versa), such as convective entrainment \citep{MA2007,Cristini2019}, convective penetration and overshooting \citep{Viallet2015}. Of these various prescriptions, it is convective penetration and overshooting that are most commonly implemented in most 1D stellar evolutionary codes \citep{Viallet2015}, and both are referred to in the literature as 'overshooting.' Both of these prescriptions are based on the traversing of convective fluid parcels beyond the limits set by the Schwarzchild or Ledoux criterion due to their inertia. In convective penetration (also known as 'step overshooting'; \citealt{Zahn1991}), the region in which these fluid particles traverse is subjected to the adiabatic temperature gradient, resulting in an extended region around the convective core being instantaneously mixed and modifying the thermal structure of the star. In convective overshooting (also known as 'diffusive exponential overshooting'; \citealt{Freytag1996,Herwig2000}), the extended region is subjected to the radiative temperature gradient, and hence only the chemical structure of star is changed. In both types of CBM regions, different functional forms of temperature gradient have been implemented, but distinguishing between them observationally is non-trivial (\citealt{Michielsen2019}, Michielsen et al. 2020, submitted). 

The net effect of overshooting is an increase in the convective core mass, and hence for the star to appear more luminous (i.e. brighter). This effect is therefore degenerate with the stellar mass, and classical observables would not be able to distinguish between both types of overshooting. However, the analyses of stars that show gravity-mode (g-mode) pulsations (e.g. \citealt{Pedersen2018,Michielsen2019}) have demonstrated that the type and amount of CBM can be constrained by modelling them asteroseismically. This has been demonstrated by a number of observational studies of g-mode pulsators (e.g. \citealt{Moravveji2015,Moravveji2016,Buysschaert2018,Walczak2019, Wu2019,Fedurco2020}).

Envelope mixing, similar to CBM, is an umbrella term that includes a number of rotational and pulsational mechanisms (see \citealt{Aerts2020} for a detailed discussion), including meridional circulation, hydrodynamical instabilities, magnetorotational instabilities and internal gravity waves. Envelope mixing in 1D stellar models is typically implemented by means of diffusive mixing coefficients in the transport equations for the regions outside of the convective core and CBM zones. There are notable exceptions, such as the advective implementation in the Geneva \citep{Ekstrom2012} and CESTAM \citep{Marques2013} stellar evolutionary codes. Due to this implementation, the envelope mixing is artifically decoupled from the CBM. It was noted by \cite{Moravveji2015, Moravveji2016} that the inclusion of envelope mixing significantly improved the fit of the g-mode pulsational frequencies and therefore they confirm that such mixing is an important component that should be included in 1D stellar models.

Due to the sensitivity of g modes to stellar interior mixing profiles, the analysis of g-mode pulsating stars allows for the calibration of these interior mixing profiles. For pulsating F-type stars, these modes are low-frequency pulsations that are excited by the convective flux-blocking mechanism  \citep{Guzik2000, Dupret2005}. At high radial order, the g modes that have the same spherical harmonic degree $\ell$ and azimuthal order $m$ (see \citealt{Aerts2010} for a complete description) have the characteristic of being equally spaced in period in the non-rotating approximation. This gives rise to the so-called asymptotic approximation of the period spacing \citep{Tassoul1980}:

\begin{equation}
\hfill \Pi_{\ell} = \frac{\Pi_{0}}{\sqrt{\ell(\ell+1)}};\hfill
\label{eq: Piell}
\end{equation}

\noindent where,

\begin{equation}
\hfill \Pi_{0} = 2\pi^{2}\left(\int_{r_{1}}^{r_{2}}N\frac{dr}{r}\right)^{-1}.\hfill
\label{eq: Pinaught}
\end{equation}

\noindent In these equations, $r$ is the distance from the stellar centre, $N$ is the Brunt-V{\"a}is{\"a}l{\"a} frequency and $r_{1}$ and $r_{2}$ are the radial boundaries of the g-mode propagation cavity in the star. Deviations from this equidistant spacing of g modes are a result of either 1) mode trapping due to the near-core chemical gradient \citep{Miglio2008}, or 2) near-core rotation, which introduces a slope into the pattern due to the pulsational periods being affected by the Coriolis force \citep{Bouabid2013}. The morphology of the observed g-mode period-spacing patterns can therefore be linked to mixing and rotational characteristics of the g mode propagation cavity (e.g. \citealt{VanReeth2015a, VanReeth2015b, VanReeth2016, Ouazzani2017, Ouazzani2020, Li2018, Li2019a, Li2020b}).

The potential for g modes in unravelling the interior mixing profiles in stars is unfortunately hampered by the highly correlated nature of the various free parameters used in the evolutionary models. \cite{Aerts2018} proposed a forward modelling scheme for the purposes of taking these correlations, as well as uncertainties due to the imperfections in the input physics of the equilibrium models, into account when performing asteroseismic modelling. In addition, restricting the ranges of the multidimensional parameter space would significantly alleviate these degeneracies in stellar modelling, resulting in more precise and accurate solutions. 

The largest number and most stringent of constraints to limit the parameter space can be obtained when studying double-lined eclipsing binary systems. One can extract the surface metallicities, effective temperatures, masses, radii, and surface gravities and rotational frequencies of the individual components. In addition to all of these parameters, binarity also enforces a co-evolutionary scenario, demanding equal ages of the individual components. These systems have been used to calibrate the amount of CBM included in stellar evolutionary models, testing either the convective penetration prescription (e.g. \citealt{Ribas2000, Torres2014, Tkachenko2014b, Stancliffe2015, Claret2016}), the convective overshooting prescription (e.g. \citealt{Tkachenko2014a, Stancliffe2015, Claret2017, Claret2018, Claret2019, Constantino2018,Johnston2019a,Johnston2019b}) or both (e.g. \citealt{Claret2017, Tkachenko2020}). A secondary advantage of studying such systems is the ability to compare three different sets of parameters: the dynamical and spectroscopic parameters, those derived from the evolutionary models, and those derived from the pulsational models.

There have been a total of 34 eclipsing binary systems with g-mode period-spacing patterns discovered so far in the \textit{Kepler} data \citep{Li2020a}, and only three have been modelled \citep{Zhang2018,Guo2019b,Zhang2020}, none of which from the viewpoint of assessing internal mixing. As such, it is the goal of our study to add to the dearth of literature on the evolutionary and asteroseismic modelling of g-mode pulsators in eclipsing binary systems. In this paper, we present the evolutionary and asterosesismic modelling of the pulsating eclipsing binary KIC9850387 that was identified as the most promising candidate in terms of g-mode asteroseismic potential in \cite{Sekaran2020}, hereafter Paper\,I. It is a short period ($P_{\text{orb}}=2.74\text{ d}$) eclipsing binary in a near-circular orbit with an intermediate-mass primary star and a solar-like secondary star. The primary star displayed remarkable $\ell=1$ and $\ell=2$ period-spacing patterns as detailed in Paper\,I. Section \ref{sec: isocloud} details the isochrone-cloud construction and fitting methodology adopted in this work, Section \ref{sec: modelling} details our asteroseismic modelling results, and we present a discussion of our results and conclusions in Section \ref{sec: conclusions}.

\begin{table}[b]
\setlength{\tabcolsep}{12pt}
\renewcommand{\arraystretch}{1.3}
\caption[Fundamental parameters of KIC9850387.]{Fundamental parameters of KIC9850387 presented in Paper\,I that were used in our evolutionary and asterosesimic modelling.}
\begin{center}
\begin{tabular}{ ccc }
 \hline
 \hline
  & \multicolumn{1}{c}{Primary} & \multicolumn{1}{c}{Secondary}\\ 
 \hline
 $M$ [M$_\odot$] & 1.66$_{-0.01}^{+0.01}$ & 1.062$_{-0.005}^{+0.003}$\\ 
$R$ [R$_\odot$] & 2.154$_{-0.004}^{+0.004}$ & 1.081$_{-0.002}^{+0.003}$\\ 
$T_{\text{eff}}$ [K] & 7335$_{-85}^{+85}$ & 6160$_{-77}^{+76}$\\ 
log~g [dex] & 3.993$_{-0.003}^{+0.003}$ & 4.396$_{-0.003}^{+0.003}$\\
$l_{\text{r}}$ & 0.893$_{-0.008}^{+0.002}$ & 0.107$_{-0.008}^{+0.002}$\\ 
{$\text{[M/H]}$ [dex]} & -0.11$_{-0.06}^{+0.06}$ & --\\
$f_{\text{rot, surf}}$ [d$^{-1}$] & 0.122$_{-0.008}^{+0.008}$ & --\\ 
 \hline
\end{tabular}
\tablefoot{\\
The errors quoted are based on 68\% HPD intervals.\\
Descriptions of the various parameters in this table:

\noindent \textbf{Primary and Secondary}
\begin{itemize}[topsep=0pt]
\item{$M$: Mass}
\item{$R$: Equivalent radius (the radius that each star would have if it was a perfect sphere)}
\item{log~g: Logarithm of the surface gravity}
\item{$T_{\text{eff}}$: Effective temperature}
\item{$l_{\text{r}}$: Light ratio of the star with respect to the total flux}
\item{$\text{[M/H]}$: Global metallicity}
\item{$f_{\text{rot, surf}}$: The surface rotational frequency of the primary star}
\end{itemize}
}
\end{center}
\label{tab: obs}
\end{table}

\section{Isochrone-cloud construction and fitting}
\label{sec: isocloud}

In Paper\,I, a full observational analysis of KIC9850387 was performed, resulting in the extraction of the orbital and fundamental parameters of the system as well as of the individual components (listed in Table \ref{tab: obs}). The $\lesssim 1$\% errors on many of these parameters are not unusual for detached eclipsing systems: The individual analyses of several tens of such systems (listed in DEBCat, \citealt{Southworth2015}) boast similar precisions. Using these parameters, we can calculate evolutionary models that best match our observations. To achieve this goal, we constructed isochrone-clouds (hereafter isoclouds) as detailed in \cite{Johnston2019b}. Each isocloud is comprised of isochrones generated from equivalent evolutionary phase (EEP) tracks (see \citealt{Dotter2016} for a full description) generated from single-star main-sequence evolutionary tracks computed with the stellar evolutionary code \textsc{mesa} (version 10348; \citealt{Paxton2011, Paxton2018}). These evolutionary tracks are generated using the same input physics as \cite{Johnston2019b}, fixing the initial helium abundance $Y_{\text{ini}}$ at 0.274 as per the cosmic B-star standard \citep{Nieva2012}, fixing $\alpha_{\text{MLT}}$ at 1.8 \citep{Joyce2018}, and using an initial metallicity ($Z_{\text{ini}}$) of 0.010 to match the spectroscopic metallicity derived for the primary star.

We adopted the Ledoux criterion to position the convective boundary and diffusive exponential overshooting as the CBM prescription according to \cite{Freytag1996} and  \cite{Herwig2000}. Restricting to one type of CBM is justified because asteroseismic analyses of a sample of A/F-type $\gamma\,$Dor pulsators have shown the results using this CBM to be indistinguishable from convective penetration \citep{Mombarg2019}. This prescription assumes an exponential decrease of the mixing efficiency with distance from the convective core as convective fluid parcels propagate further away from the core into the radiative zone. The amount of diffusive mixing in the CBM region [$D_{\text{CBM}}(r)$] is calculated using the following equation:

\begin{equation}
\hfill D_{\text{CBM}}(r) = D_{\text{0}} \ \text{exp}\left[\frac{-2(r-r_{\text{0}})}{f_{\text{ov}}H_{\text{p}}}\right].\hfill
\label{eq: CBM}
\end{equation}

\noindent Here, the reference radius $r_{0}=r_{\text{cc}}-f_{0}H_{\text{p}}$, where $r_{\text{cc}}$ is the radius of the convective core according to the Ledoux criterion, ${f_{\text{0}}=0.002}$ and $H_{\text{p}}$ is the local pressure scale height. $D_{\text{0}}$ is the amount of diffusive mixing within the inner edge of the convective-core boundary at the reference radius $r_{0}$ and $f_{\text{ov}}$ is the length scale over which the diffusive exponential overshooting applies. 

\begin{table}[t]
\setlength{\tabcolsep}{12pt}
\renewcommand{\arraystretch}{1.3}
\caption[Ranges of $M_{\text{ini}}$, $f_{\text{ov}}$ and log~$D_{\text{mix}}$ used to create evolutionary tracks.]{Ranges of $M_{\text{ini}}$, $f_{\text{ov}}$ and log~$D_{\text{mix}}$ used to create evolutionary tracks.}
\begin{center}
\begin{tabular}{ cccc } 
 \hline
 \hline
  & Min & Max & Step\\ 
 \hline
$M_\text{ini}$ [$M_{\odot}$] & 0.80 & 2.00 & 0.05\\
$f_{\text{ov}}$ & 0.005 & 0.040 & 0.005\\
log~$D_{\text{mix}}$ & 0.0 & 4.0 & 0.5\\
 \hline
\end{tabular}
\tablefoot{$f_{\text{ov}}$ and $D_{\text{mix}}$ are fixed at 0 for evolutionary tracks with $M_\text{ini}<1.15$.}
\end{center}
\label{tab: mesa_grid}
\end{table}

The prescriptions of \cite{Johnston2019b} include ranges for $f_{\text{ov}}$ and the amount of diffusive envelope mixing $D_{\text{mix}}$, taken from the asteroseismic calibrations of \cite{Aerts2015,Bowman2020} and \cite{Moravveji2015,Moravveji2016} respectively. More specifically, the diffusive envelope mixing adopted in \cite{Johnston2019b} takes the form of $D_{\text{mix}}\propto\rho^{-1/2}$, which is an approximation of the mixing due to internal gravity waves taken from 2D hydrodynamical simulations \citep{Rogers2017} and implemented in \textsc{mesa} by \cite{Pedersen2018}. However, these parameters have been calibrated using g-mode pulsators, and are therefore only appropriate for stars that develop a substantial convective core over the main sequence ($M_{\text{ini}}\gtrsim 1.15$ M$_\odot$ as noted by \citealt{Dotter2016}). The low dynamical mass of the secondary (${M_{\text{s}}=1.0617_{-0.005}^{+0.003}}$~M$_\odot$) necessitates the calculation of evolutionary tracks of stars below this limit. \cite{Choi2016} computed a grid of solar-calibrated stellar models and isochrones but only include rotational mixing in their tracks with $M_{\text{ini}}\geq 1.15\text{ M}_{\odot}$ to reflect the slow rotational frequencies observed in low-mass stars. Considering that rotational mixing is the only form of envelope mixing present in their tracks, it can be concluded that they effectively set the diffusive envelope mixing at zero for their evolutionary tracks with $M_{\text{ini}}<1.15\text{ M}_{\odot}$. In addition, a significant portion of the envelope of low-mass stars is convective, implying that mixing in the radiative region outside of the core would have a minimal impact on their evolution. We therefore fix both $f_{\text{ov}}$ and $D_{\text{mix}}$ at 0 for our evolutionary tracks for $M_{\text{ini}}<1.15$~M$_\odot$. The ranges of $M_{\text{ini}}$, $f_{\text{ov}}$ and log~$D_{\text{mix}}$ used to create our evolutionary tracks are listed in Table \ref{tab: mesa_grid}. It should be noted that our use of an upper bound of 4.0 for log~$D_{\text{mix}}$ is atypically high for intermediate-mass g-mode pulsators (e.g. \citealt{Mombarg2019} had used an upper bound of 1.0 for log~$D_{\text{mix}}$). Such high values allow for the testing of the existence of tidal mixing mechanisms in the stellar interior.

An isochrone is traditionally constructed from evolutionary model tracks with the same input physics and the same free-parameter (e.g. $f_{\text{ov}}$ and $D_{\text{mix}}$) values. However, due to our choice of fixing $f_{\text{ov}}$ and log $D_{\text{mix}}$ at 0 for $M_{\text{ini}}<1.15$~M$_\odot$, we effectively have to fuse the isochrones created from the $M_{\text{ini}}\leq1.10$~M$_\odot$ and $M_{\text{ini}}\geq1.15$~M$_\odot$ evolutionary tracks together: A single isochrone therefore has an $f_{\text{ov}}$ and log $D_{\text{mix}}$ value of 0 for $0.80\text{ M}_{\odot}\leq M_{\text{ini}}\leq1.10\text{ M}_{\odot}$ and a fixed combination of $f_{\text{ov}}$ and log $D_{\text{mix}}$ as per the ranges listed in Table \ref{tab: mesa_grid} for $1.15\text{ M}_{\odot}\leq M_{\text{ini}}\leq2.00\text{ M}_{\odot}$, with a total of 1000 datapoints with $0.80\leq M_{\text{ini}}\leq2.00$. Similarly, each isocloud, which is a combination of all isochrones at an age $\tau$ for all combinations of $f_{\text{ov}}$ and log~$D_{\text{mix}}$ in our grid, exhibits the same behaviour. Each of our isoclouds comprise a total of 72~000 datapoints: 8 $f_{\text{ov}}$ values $\times$ 9 $\log{D_{\text{mix}}}$ values $\times$ 1000 datapoints per isochrone.

\begin{table}[b]
\setlength{\tabcolsep}{12pt}
\renewcommand{\arraystretch}{1.3}
\caption[Evolutionary model parameters of KIC9850387.]{Evolutionary model parameters of KIC9850387 extracted from isochrone-cloud fitting, based on the intersection of the dynamical $T_{\text{eff}}$ and log~g constraints with the isocloud parameters corresponding to the 95\% HPD interval of the Monte-Carlo age distribution ($\tau_{\text{MC}}=1.3^{+1.5}_{-0.2}$ Gyr).}
\begin{center}
\begin{tabular}{ ccc }
 \hline
 \hline
  & \multicolumn{1}{c}{Primary} & \multicolumn{1}{c}{Secondary}\\ 
 \hline
 $\tau$ [Gyr] & $2.0\pm0.8$ & $1.2\pm0.5$\\
 $M$ [M$_\odot$] & $1.6\pm0.1$ & $1.07\pm0.02$\\ 
$X_{\text{c}}$ & $0.3\pm0.1$ & $0.57\pm0.05$\\ 
$f_{\text{ov}}$ & $0.02\pm0.02$ & --\\
log $D_{\text{mix}}$ & $1.5\pm1.5$ & --\\ 
$M_{\text{cc}}$ [M$_\odot$] & $0.17\pm0.03$ & --\\
 \hline
\end{tabular}
\end{center}
\label{tab: evol_params}
\end{table}

\begin{figure*}[!htp]
\includegraphics[width=\hsize]{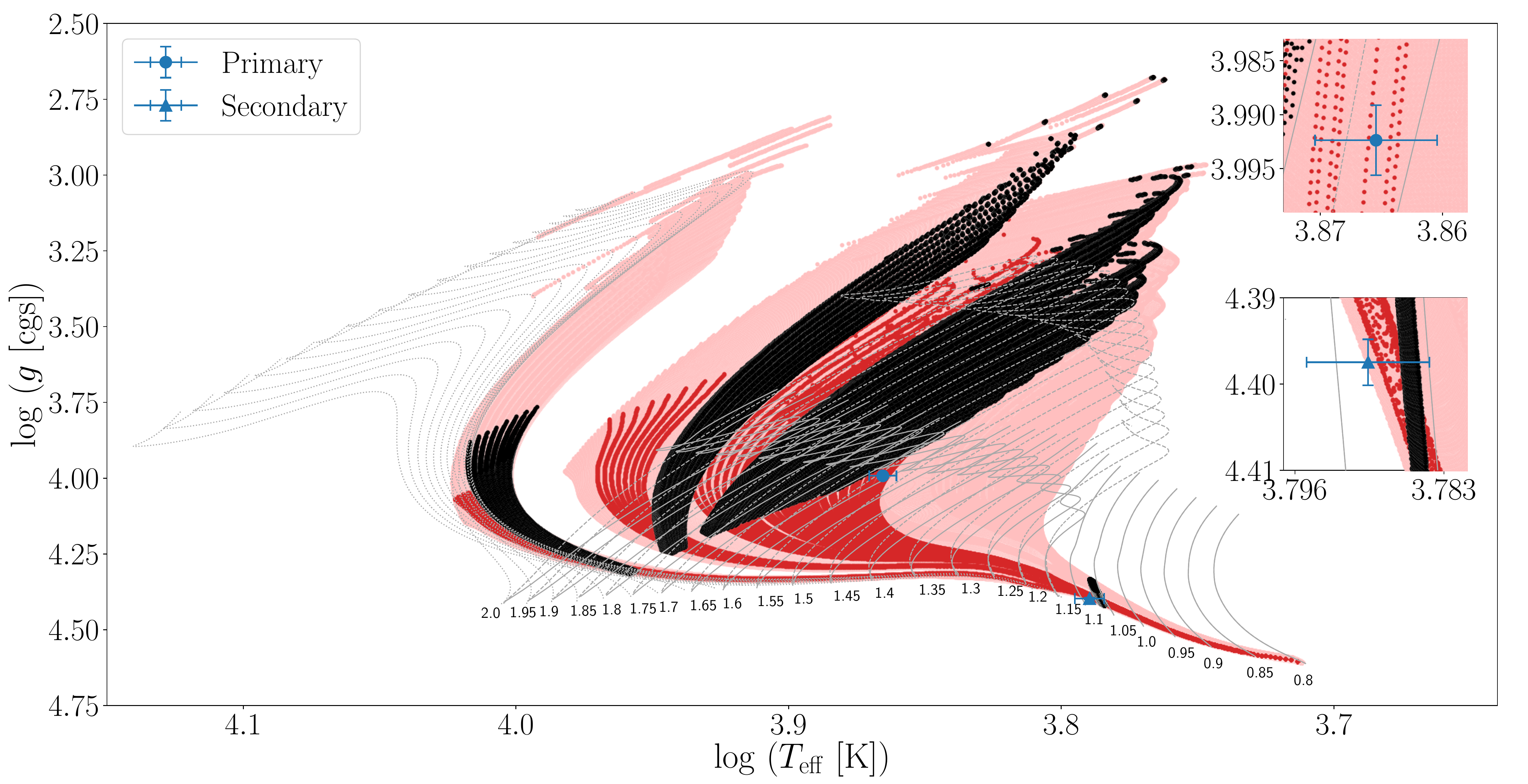}
\caption[Isocloud fit to the components of KIC9850387 on the $\text{log }T_{\text{eff}}-\text{log }g$ diagram.]{Isocloud fit to the components of KIC9850387 on the $\text{log }T_{\text{eff}}-\text{log }g$ diagram. The red regions correspond to the best-fitting isocloud, and the pink regions represent the isoclouds in the 95\% HPD interval of the fit. The black regions on the isocloud represent the $\text{log }T_{\text{eff}}-\text{log }g$ values corresponding to the dynamical masses of the individual components. The grey curves are \textsc{mesa} evolutionary tracks with $f_{\text{ov}}=0.005$ and $\text{log }D_{\text{mix}}=0.0$ (solid), $f_{\text{ov}}=0.04$ and $\text{log }D_{\text{mix}}=0.0$ (dashed), and $f_{\text{ov}}=0.04$ and $\text{log }D_{\text{mix}}=4.0$ (dotted) with their corresponding masses (in units of M$_{\odot}$) indicated at the ZAMS of each track. The inset plots are magnified regions around the position of the primary (top) and secondary (bottom) component.}
\label{fig: isocloud_fit}
\end{figure*}

In order to obtain the best-fitting isocloud to the dynamical parameters of KIC9850387, we perturbed the primary and secondary $T_{\text{eff}}$ and log~g within their 1$\sigma$ observational errors in a Monte-Carlo framework for 1000 iterations, and retained the best-fitting isocloud in each iteration. Due to the fact that an isocloud is effectively a set of curves, we calculated the reduced $\chi^{2}$ ($\chi^{2}_{\text{red}}$) value for each of the 72~000 datapoints in an isocloud with respect to the primary and secondary $T_{\text{eff}}$ and log~g (i.e. 144~000 total $\chi^{2}_{\text{red}}$ values), and took the sum of the 50 smallest $\chi^{2}_{\text{red}}$ values for both the primary and secondary star (i.e. the sum of 100 $\chi^{2}_{\text{red}}$ values across both components) as our goodness-of-fit metric. In addition, to account for the fact that the stellar models are in much better agreement with dynamical parameters for low-mass stars (e.g. \citealt{Chaplin2014,Higl2017}), we chose to bias the isocloud fitting towards the secondary component. This was done by decreasing the errors on the secondary parameters by a factor of five, effectively weighting the secondary parameters five times more heavily than the primary ones. We then sample this distribution of the 1000 best-fitting isoclouds and determined the evolutionary parameters ($\tau$, $M$, $f_{\text{ov}}$, log $D_{\text{mix}}$, the central hydrogen fraction $X_{\text{c}}$, and the inferred convective-core mass $M_{\text{cc}}$) based on the intersection of the dynamical $T_{\text{eff}}$\footnote{We refer to the $T_{\text{eff}}$ determined spectroscopically by fixing the log~g at the dynamical value, as detailed in Section 3.3 of Paper\,I, as the 'dynamical' $T_{\text{eff}}$.}  and log~g constraints with the isocloud parameters corresponding to the 95\% highest posterior density (HPD) interval\footnote{It should be noted that we use a 95\% HPD only for our isocloud fit based on model-independent dynamical constraints, but 68\% confidence and HPD intervals for the model-dependent evolutionary and asteroseismic parameters presented in the rest of this paper. This was done to ensure sufficient baseline coverage of the parameter space of the stellar structural models.} of the Monte-Carlo age distribution, corresponding to $\tau_{\text{MC}}=1.3^{+1.5}_{-0.2}$ Gyr. These parameters are listed in Table \ref{tab: evol_params}, and the $\text{log }T_{\text{eff}}-\text{log }g$ (or Kiel) diagram displaying our isocloud fitting results is shown in Figure \ref{fig: isocloud_fit}. The large asymmetry in the errors of $\tau_{\text{MC}}$ is a consequence of older isoclouds covering a larger fraction of the $T_{\text{eff}}-\log{g}$ parameter space at higher masses, as the varying amounts of core-boundary and envelope mixing included in the models cause the corresponding evolutionary tracks to diverge more strongly with increasing age. These isoclouds therefore start to overlap more greatly at older ages, allowing for a greater range in the upper age bound. It should be noted that the isocloud diverges into three distinct sub-clouds due to the logarithmic scaling of the $\text{log }D_{\text{mix}}$ values in our grid.

As noted in Paper\,I, there is a slight discrepancy between the evolutionary and dynamical masses of the primary star. However, our isoclouds were calculated from evolutionary tracks using a fixed $Z_{\text{ini}}=0.010$, which is close to the best-fitting spectroscopic surface metallicity but its associated errors are not taken into account. Varying the initial metallicity within the spectroscopic errors (i.e. $0.008 \lesssim Z_{\text{ini}} \lesssim 0.012$) resolves this discrepancy, but introducing $Z_{\text{ini}}$ as another free parameter leads to additional degeneracies (e.g. between $Z_{\text{ini}}$ and $M$, and between $Z_{\text{ini}}$ and $f_{\text{ov}}$). These degeneracies would have further propagated to our subsequent asteroseismic analyses, and as such we chose not to include $Z_{\text{ini}}$ as a free parameter for our study.

It can be seen in Table \ref{tab: evol_params} that both $f_{\text{ov}}$ and $D_{\text{mix}}$ are unconstrained in our fit, although $\text{log }D_{\text{mix}}>3$ is incompatible with the position of the star in the Kiel diagram (cf. Figure \ref{fig: isocloud_fit}). This has been noted in other studies as well (e.g. \citealt{Johnston2019a, Johnston2019b, Mombarg2019, Tkachenko2020}). However, as noted by those very same studies, $M_{\text{cc}}$ is well constrained, and our value of $M_{\text{cc}}/M_{*}=0.11\pm0.02$ is compatible with the range (${0.075 \lesssim M_{\text{cc}}/M_{*} \lesssim 0.125}$) reported for a sample of $\gamma\,$Dor stars by \cite{Mombarg2019}.

\begin{figure}[!hb]
\includegraphics[width=\hsize]{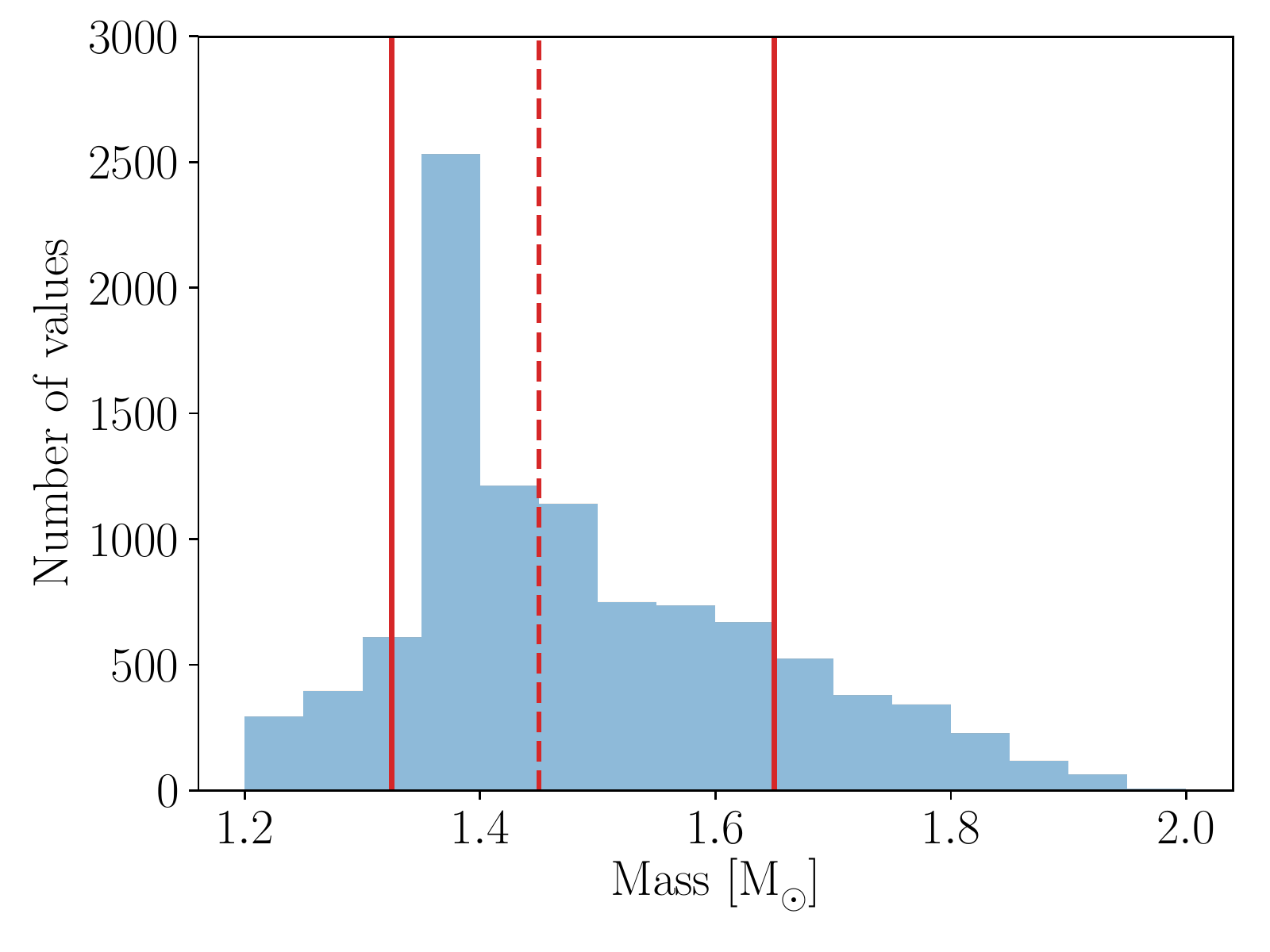}
\caption[Example of the distribution of $M$ values in the posterior distribution selected through our Monte-Carlo fitting framework.]{Example of the distribution of $M$ values in the posterior distribution of models selected through our Monte-Carlo fitting framework. The vertical dashed line represents the median value and the vertical solid lines represent the upper and lower bounds of 68\% HPD of the $M$ values.}
\label{fig: M_hist}
\end{figure}

\begin{figure*}[!htp]
\includegraphics[width=\hsize]{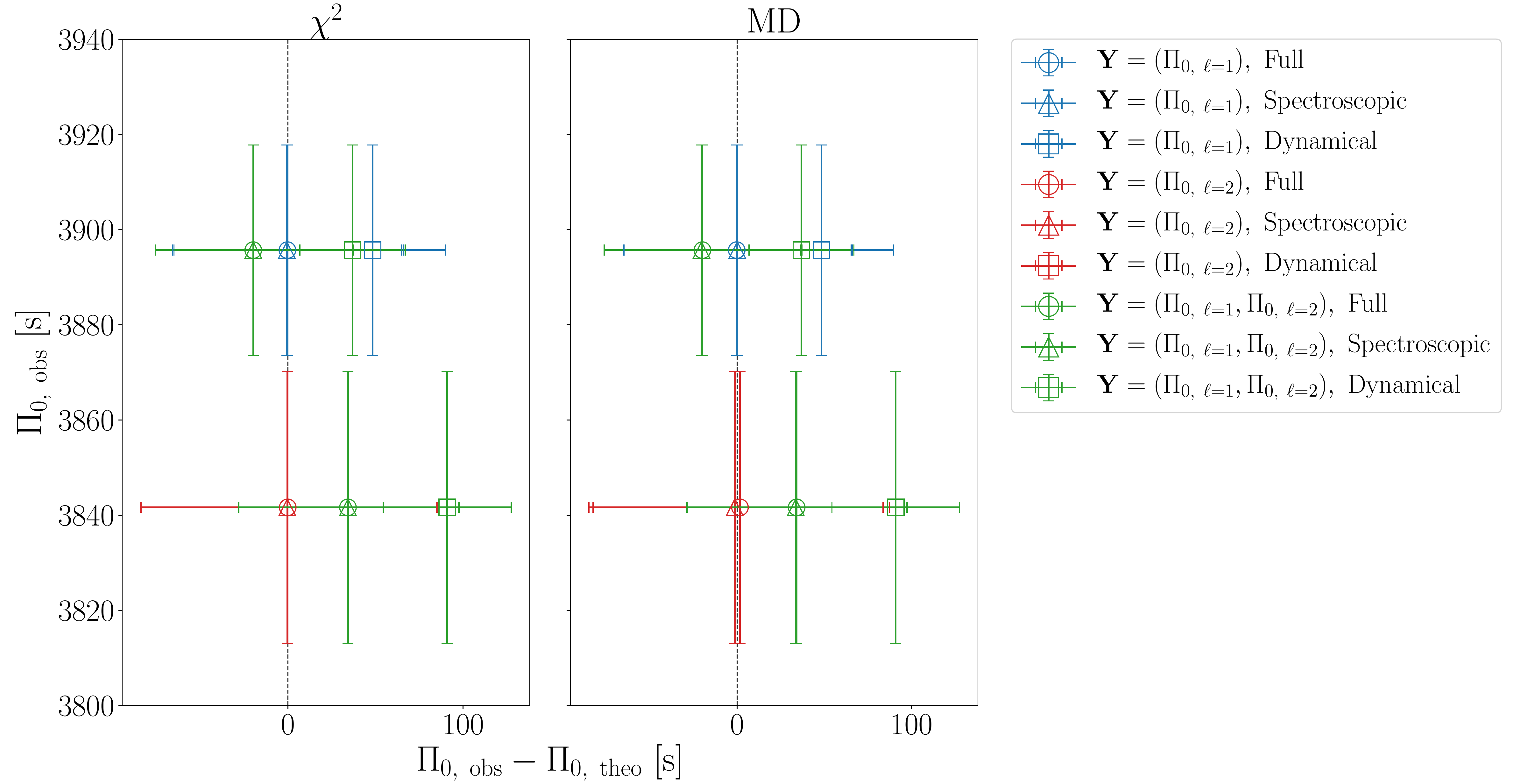}
\caption[Best-fitting $\Pi_{0}$ values of our $\Pi_{0}$-based asteroseismic modelling.]{Best-fitting $\Pi_{0}$ values of our $\Pi_{0}$-based asteroseismic modelling, based on three different grid setups (full, spectroscopic and dynamical). The left panel shows the results of using a $\chi^{2}_{\text{red}}$ merit function and the right panel shows the results of using an MD merit function. The vertical dashed black line represents the zero-point of the difference between the observational and theoretical $\Pi_{0}$ (i.e. $\Pi_{0,\text{ obs}}-\Pi_{0,\text{ theo}}=0$). The error bars are based on 68\% HPD intervals of the Monte-Carlo parameter distributions.}
\label{fig: Pi0_results}
\end{figure*}

\section{Asteroseismic modelling}
\label{sec: modelling}

Our asteroseismic modelling is based on two different sets of asteroseismic observables: the asymptotic period spacing $\Pi_{0}$ (described in Eqs. (\ref{eq: Piell}) and (\ref{eq: Pinaught}) and the individual period-spacing values ($\Delta P_{\ell}$) of the $\ell=1$ and $\ell=2$ modes. As noted by \cite{Zhang2020} and \cite{Li2020b}, the primary star of KIC9850387 is an extremely slow rotator, with \cite{Li2020b} reporting a near-core rotational frequency of $f_{\text{rot, core}}=0.0053$ d$^{-1}$. We are therefore able to estimate the observational values of $\Pi_{0}$ from ${\Pi_{0\text{, obs}}\simeq\overline{\Delta P_{\ell}}*\sqrt{\ell(\ell+1)}}$ in the approximation of a non-rotating star. The determination of $\Pi_{0}$
is more complicated for fast rotators (e.g. \citealt{VanReeth2016, VanReeth2018, Li2018, Li2019a, Li2019b, Li2020a, Li2020b}) as this requires an estimation of the slope of the period spacing pattern, which is absent for KIC9850387. We are therefore able to perform asteroseismic modelling using the quantity $\Pi_{0}$ directly, without estimation of the rotation frequency, through Eqs. (\ref{eq: Piell}) and (\ref{eq: Pinaught}), by comparing our ${\Pi_{0\text{, obs}}}$ estimates with values predicted from the non-rotating stellar evolutionary models. One of the advantages of isocloud fitting (see Section \ref{sec: isocloud}) is that it provides age constraints for our stellar evolutionary models. We therefore restricted our evolutionary models to the ages that were within the 95\% HPD interval of the Monte Carlo isocloud-fitting age distribution ($\tau_{\text{MC}}=1.3^{+1.5}_{-0.2}$ Gyr). 

In order to obtain theoretical $\Delta P_{\ell}$, values, one would have to calculate pulsational models. These models require stellar structural models (i.e. a model describing the temperature, density, chemical and mixing profiles from the centre to the surface of the star) as an input. Using our isocloud-fitting age constraint, we computed a grid of stellar structural models with masses between 1.2~M$_{\odot}$ to 2.0~M$_{\odot}$ in steps of 0.02 in $X_{\text{c}}$ as an input to compute pulsational mode predictions in the adiabatic framework using the stellar pulsation code \textsc{gyre} (revision 5.2; \citealt{Townsend2013,Townsend2018}). We then use the theoretical pulsational frequencies of the zonal $\ell=1$ and $\ell=2$ modes to construct theoretical period-spacing patterns and confront these to the observed ones, which is a more detailed analysis compared to just using 
$\Pi_{0}$.

For both approaches, our fitting methodology involves the perturbation of the $\Pi_{0\text{, obs}}$ and observational $\Delta P_{\ell}$ values of the period-spacing patterns ($\Delta P_{\ell\text{, obs}}$) within their respective 1$\sigma$ observational errors in a Monte-Carlo framework for 10000 iterations, and then retaining the parameters of the best-fitting model in each iteration. To determine the best-fitting model within this framework, we used two different merit functions: 1) the $\chi^{2}_{\text{red}}$ and 2) the Mahalanobis Distance (MD). As discussed in detail in \cite{Aerts2018}, the MD is a maximum-likelihood point estimator that takes into account uncertainties in the theoretical asteroseismic predictions and that 
treats correlations in the free parameters of our stellar model grid appropriately by incorporating its covariance matrix (\textbf{V}) into the distance calculation:

\begin{equation}
\hfill \text{MD} = (\textbf{Y}_{\text{theo}}-\textbf{Y}_{\text{obs}})^{\text{T}}(\textbf{V}+\boldsymbol{\Lambda})^{-1}(\textbf{Y}_{\text{theo}}-\textbf{Y}_{\text{obs}}).\hfill
\label{eq: MD}
\end{equation}

\noindent In this equation, $\textbf{Y}_{\text{theo}}$ and $\textbf{Y}_{\text{obs}}$ are the vectors representing the theoretical asteroseismic predictions and asteroseismic observables that are being compared, and the matrix $\boldsymbol{\Lambda}$ is a diagonal matrix with squared observational errors as each of the diagonal elements. 
The free parameters to be estimated, determining the theoretical asteroseismic predictions, are $M$, $X_c$, $f_{\rm ov}$ and $\log D_{\rm mix}$.
The MD takes uncertainties in the theoretical asteroseismic predictions due to the imperfect input physics of the stellar structural models in the grid into account. It does so via the matrix $\textbf{V}$, where we assume that its components are well described by the variance induced by the whole range of the four free parameters defining the theoretical asteroseismic predictions.

\subsection{$\Pi_{0}$ modelling results}
\label{subsec: Piell}

\begin{figure*}[!htp]
\includegraphics[width=\hsize]{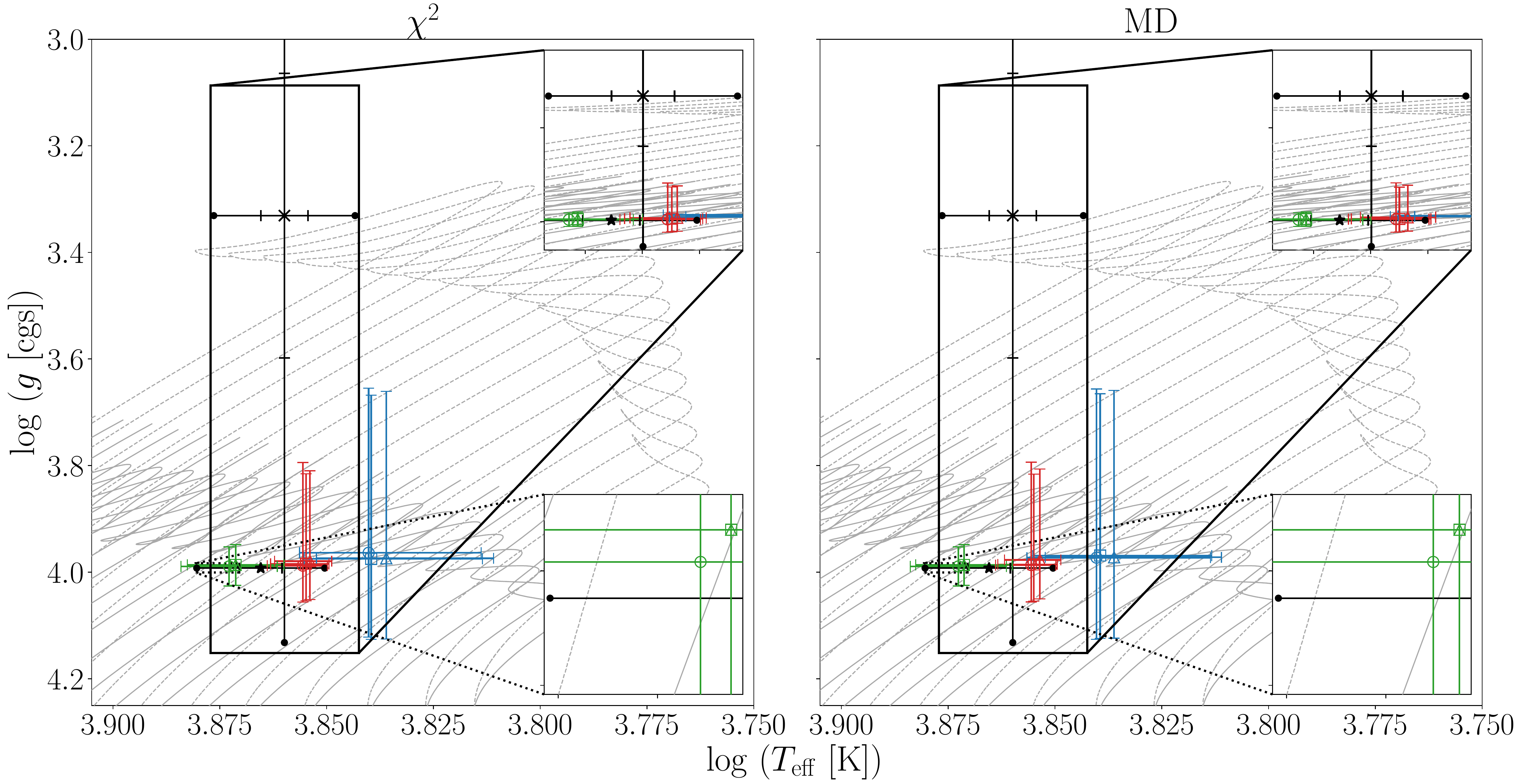}
\caption[Positions of the best-fitting models from $\Pi_{0}$-based asteroseismic modelling on Kiel diagrams.]{Positions of the best-fitting models from $\Pi_{0}$-based asteroseismic modelling on Kiel diagrams. The main plots show the results of using the whole grid in the fit, while the inset plots show the results of applying the `pseudo-single-star' spectroscopic $T_{\text{eff}}$ and log~g (solid box) and dynamical $M$, $T_{\text{eff}}$ and log~g (dotted box) constraints. The observed positions of the star according to the spectroscopic and dynamical parameters are represented by black 'X' symbols and stars respectively. The 1$\sigma$ and 3$\sigma$ spectroscopic and dynamical error bars are represented with straight end-caps and ball end-caps respectively. The ${\bf{Y_{\rm obs}}=(\Pi_{0, \ \boldsymbol{\ell}=1})}$, ${\bf{Y_{\rm obs}}=(\Pi_{0, \ \boldsymbol{\ell}=1})}$ and ${\bf{Y_{\rm obs}}=(\Pi_{0, \ \boldsymbol{\ell}=1}, \Pi_{0, \ \boldsymbol{\ell}=2})}$ solutions for the different grid subsets are represented by the same colours and symbols as in Figure \ref{fig: Pi0_results}. The left panel shows the results of using a $\chi^{2}_{\text{red}}$ merit function and the right panel shows the results of using an MD merit function. The error bars on the asteroseismic solutions are based on 68\% HPD intervals of the Monte Carlo parameter distributions. The solid and dashed grey curves are the same as in Figure \ref{fig: isocloud_fit}.}
\label{fig: Pi0_Kiel}
\end{figure*}

We performed our $\Pi_{0}$ modelling by restricting the size of our grid based on varying types of observational constraints. Due to the tight constraints imposed by the dynamical parameters, a grid setup created by restricting the parameter space based on the dynamical $M$, $T_{\text{eff}}$ and log~g would necessarily be very small, comprising only a few tens of models in total. In order to investigate the effects of different levels of parameter space restriction on our results, we chose to construct a third grid setup of intermediate detail and size between the full evolutionary grid and the small asteroseismic grid based on the dynamical constraints. To that end, we obtained a second set of $T_{\text{eff}}$ and log~g constraints by performing atmospheric parameter determination for the primary star of KIC9850387 (as detailed in Section 3.3 of Paper\,I) but ignoring any binary or eclipse information (i.e. treating the system as if it was a single star and performing a classical atmospheric analysis).

Using these `pseudo-single-star' spectroscopic and dynamical parameters, we created three grid setups: 1) The full grid, as the name suggests, is the grid resulting from the 95\% HPD interval of the Monte-Carlo age distribution of our isocloud fit (as described in Section \ref{sec: isocloud}) without any other constraints; 2) The spectroscopic grid is a subset of the full grid based on the 3$\sigma$ intervals of the `pseudo-single-star' spectroscopic $T_{\text{eff}}$ and log~g values; and 3) the dynamical grid is a subset of the full grid based on the 3$\sigma$ intervals of the dynamical $M$, $T_{\text{eff}}$ and log~g values. The models in the spectroscopic and dynamical grids have $6965\text{ K}\leq T_{\text{eff}}\leq 7520\text{ K}$ and $3.09\leq\text{log }g\leq4.13$ (spectroscopic grid), and $1.63\text{ M}_{\odot}\leq M_{\text{ini}}\leq 1.70\text{ M}_{\odot}$, $7081\text{ K}\leq T_{\text{eff}}\leq 7590\text{ K}$ and $3.9826\leq\text{log }g\leq4.0021$ (dynamical grid) respectively. It should be noted that we use 3$\sigma$ intervals to construct our grids to ensure that the grids are sufficiently large and to introduce sufficient variance into the grids, which as mentioned significantly impacts our MD calculations.

For each grid setup, we fitted the $\Pi_{0, \ \text{obs}}$ estimates for each of the dipole ($\ell=1$) and quadrupole ($\ell=2$) modes separately, as well as combined into a single vector with two components, resulting in three different solutions. More specifically, for each of our grid setups (1,2 and 3), we used three different sets of observables and predictions in independent fits: a) $\bf{Y}=(\Pi_{0, \ \boldsymbol{\ell}=1})$, b) $\bf{Y}=(\Pi_{0, \ \boldsymbol{\ell}=2})$, and c) $\bf{Y}=(\Pi_{0, \ \boldsymbol{\ell}=1}, \Pi_{0, \ \boldsymbol{\ell}=2})$. The results for each parameter setup using either the $\chi^{2}_{\text{red}}$ or the MD merit functions are listed in Table \ref{tab: Pi0_as} (full grid), Table \ref{tab: Pi0_as_spec} (spectroscopic grid) and Table \ref{tab: Pi0_as_dyn} (dynamical grid). Comparisons of the $\chi^{2}_{\text{red}}$ and the MD results in terms of the best-fitting $\Pi_{0}$ values and positions on Kiel diagrams are shown in Figures \ref{fig: Pi0_results} and \ref{fig: Pi0_Kiel} respectively. The errors are based on 68\% HPD intervals of the Monte-Carlo parameter distributions (e.g. see Figure \ref{fig: M_hist}). This allows for proper comparison with the evolutionary modelling and dynamical results, which are based on 68\% confidence intervals. 

It should be noted that in some cases, the models that were selected in every single Monte-Carlo iteration had the same values for the free parameters, due to the limited resolution of the grid and its subsets. In these cases, we adopt upper- and lower-bound errors on the estimated parameters ($M$, $X_{\text{c}}$, $f_{\rm ov}$ and $\log D_{\rm mix}$) that are of a single step size in the grid (i.e. 0.5~$M_{\odot}$ in $M$, 0.005 in $f_{\text{ov}}$ and 0.5 in $\log{D_{\text{mix}}}$) and propagate these errors to the inferred parameters ($R$, $T_{\text{eff}}$, log~g, $M_{\text{cc}}$ and $\tau$). These errors (hereafter single-grid-step errors) should be interpreted as a conservative upper limit on the precision of the extracted parameters based on the resolution of our grid. Parameters for which these single-grid-step errors are assumed are indicated by $^{*}$ next to the values in the tables.

\begin{table*}[!htp]
\setlength{\tabcolsep}{6pt}
\renewcommand{\arraystretch}{1.3}
\caption[Results of asteroseismic modelling with ${\textbf{Y}=(\bf{\Pi_{0}})}$ using the full grid of pulsational models.]{Results of asteroseismic modelling with ${\textbf{Y}=(\bf{\Pi_{0})}}$ using the full grid of pulsational models, with either a $\chi^{2}_{\text{red}}$ or MD merit function.}
\begin{center}
{\small
\begin{tabular}{ c|ccc|ccc } 
 \hline
 \hline
 & \multicolumn{3}{c|}{$\chi^{2}_{\text{red}}$} & \multicolumn{3}{c}{$\text{MD}$}\\
 \hline
  & $\bf{Y_{\rm obs}}=(\Pi_{0, \ \boldsymbol{\ell}=1})$ & $\bf{Y_{\rm obs}}=(\Pi_{0, \ \boldsymbol{\ell}=2})$  & $\bf{Y_{\rm obs}}=(\Pi_{0, \ \boldsymbol{\ell}=1}, \Pi_{0, \ \boldsymbol{\ell}=2})$ & $\bf{Y_{\rm obs}}=(\Pi_{0, \ \boldsymbol{\ell}=1})$ & $\bf{Y_{\rm obs}}=(\Pi_{0, \ \boldsymbol{\ell}=2})$  & $\bf{Y_{\rm obs}}=(\Pi_{0, \ \boldsymbol{\ell}=1}, \Pi_{0, \ \boldsymbol{\ell}=2})$ \\ 
 \hline
$M$ [$\text{M}_{\odot}$] & $1.5_{-0.1}^{+0.2}$ & $1.5_{-0.1}^{+0.2}$ & $1.5_{-0.1}^{+0.2}$ & $1.5_{-0.1}^{+0.2}$  & $1.5_{-0.1}^{+0.2}$ & $1.5_{-0.1}^{+0.2}$\\
$f_{\text{ov}}$ & $0.025_{-0.010}^{+0.008}$ & $0.025_{-0.010}^{+0.008}$ & $0.025_{-0.010}^{+0.008}$ & $0.025_{-0.010}^{+0.008}$ & $0.025_{-0.010}^{+0.008}$ & $0.025_{-0.010}^{+0.008}$\\
log $D_{\text{mix}}$ & $1.5_{-0.8}^{+0.8}$ & $1.5_{-0.8}^{+0.8}$ & $1.5_{-0.8}^{+0.8}$ & $1.5_{-0.8}^{+0.8}$ & $1.5_{-0.8}^{+0.8}$ & $1.5_{-0.8}^{+0.8}$\\
$X_{\text{c}}$ & $0.28_{-0.12}^{+0.16}$ & $0.28_{-0.12}^{+0.15}$ & $0.28_{-0.14}^{+0.15}$ & $0.28_{-0.12}^{+0.16}$ & $0.26_{-0.12}^{+0.16}$ & $0.26_{-0.11}^{+0.17}$\\
\hline
$R$ [$\text{R}_{\odot}$] & $2.0_{-0.4}^{+1.3}$ & $2.0_{-0.3}^{+1.3}$ & $2.0_{-0.3}^{+1.5}$ & $2.0_{-0.4}^{+1.3}$ & $2.0_{-0.3}^{+1.4}$ & $2.0_{-0.3}^{+1.4}$\\
$T_{\text{eff}}$ [K] & $6934_{-445}^{+253}$ & $6866_{-419}^{+245}$ & $6913_{-423}^{+259}$ & $6934_{-442}^{+262}$ & $6864_{-412}^{+247}$ & $6906_{-420}^{+262}$\\
log~g [dex] & $4.0_{-0.3}^{+0.2}$ & $4.0_{-0.3}^{+0.1}$ & $4.1_{-0.3}^{+0.1}$ & $4.0_{-0.3}^{+0.2}$ & $4.0_{-0.3}^{+0.1}$ & $4.1_{-0.3}^{+0.1}$\\
$M_{\text{cc}}$ [$M_{\odot}$] & $0.145_{-0.034}^{+0.009}$ & $0.140_{-0.025}^{+0.009}$ & $0.14_{-0.04}^{+0.01}$ & $0.144_{-0.035}^{+0.009}$ & $0.140_{-0.025}^{+0.009}$ & $0.143_{-0.038}^{+0.010}$\\
$\tau$ [Gyr] & $1.7_{-0.3}^{+0.5}$ & $1.8_{-0.3}^{+0.4}$ & $1.8_{-0.3}^{+0.5}$ & $1.7_{-0.4}^{+0.5}$ & $1.8_{-0.3}^{+0.4}$ & $1.8_{-0.3}^{+0.5}$\\
 \hline
$\chi^{2}_{\text{red, min}}|\text{MD}_{\text{min}}$ & $2.09\cdot 10^{-14}$ & $8.35\cdot 10^{-16}$ & $7.58\cdot 10^{-7}$ & $1.70\cdot 10^{-19}$ & $4.23\cdot 10^{-16}$ & $7.27\cdot 10^{-8}$\\ 
 \hline
\end{tabular}
}
\tablefoot{The table is divided into two, with the estimated parameters ($M$, $X_{\text{c}}$, $f_{\rm ov}$ and $\log D_{\rm mix}$) presented in the top four rows and the inferred parameters ($R$, $T_{\text{eff}}$, log~g, $M_{\text{cc}}$ and $\tau$) in the bottom five rows. The errors quoted are based on 68\% HPD intervals of the Monte Carlo parameter distributions. The minimum values of each merit function across all of the Monte Carlo iterations for each parameter setup are quoted in the bottom row of the table.}
\end{center}
\label{tab: Pi0_as}

\vspace{6pt}

\setlength{\tabcolsep}{6pt}
\renewcommand{\arraystretch}{1.3}
\caption[Same as Table \ref{tab: Pi0_as}, but for a grid subset based on the 3$\sigma$ interval of the `pseudo-single-star' spectroscopic $T_{\text{eff}}$ and log~g values.]{Same as Table \ref{tab: Pi0_as}, but for a grid subset based on the 3$\sigma$ interval of the `pseudo-single-star' spectroscopic $T_{\text{eff}}$ and log~g values (see footnote).}
\begin{center}
{\small
\begin{tabular}{ c|ccc|ccc }
 \hline
 \hline
 & \multicolumn{3}{c|}{$\chi^{2}_{\text{red}}$} & \multicolumn{3}{c}{$\text{MD}$}\\
 \hline
  & $\bf{Y_{\rm obs}}=(\Pi_{0, \ \boldsymbol{\ell}=1})$ & $\bf{Y_{\rm obs}}=(\Pi_{0, \ \boldsymbol{\ell}=2})$  & $\bf{Y_{\rm obs}}=(\Pi_{0, \ \boldsymbol{\ell}=1}, \Pi_{0, \ \boldsymbol{\ell}=2})$ & $\bf{Y_{\rm obs}}=(\Pi_{0, \ \boldsymbol{\ell}=1})$ & $\bf{Y_{\rm obs}}=(\Pi_{0, \ \boldsymbol{\ell}=2})$  & $\bf{Y_{\rm obs}}=(\Pi_{0, \ \boldsymbol{\ell}=1}, \Pi_{0, \ \boldsymbol{\ell}=2})$\\ 
 \hline
$M$ [$\text{M}_{\odot}$] & $1.55_{-0.08}^{+0.15}$ & $1.55_{-0.08}^{+0.18}$ & $1.55_{-0.10}^{+0.15}$ & $1.55_{-0.08}^{+0.15}$  & $1.55_{-0.08}^{+0.18}$ & $1.55_{-0.10}^{+0.15}$ \\
$f_{\text{ov}}$ & $0.010_{-0.003}^{+0.008}$ & $0.010_{-0.003}^{+0.005}$ & $0.010_{-0.003}^{+0.008}$ & $0.010_{-0.003}^{+0.008}$ & $0.010_{-0.003}^{+0.005}$ & $0.010_{-0.003}^{+0.008}$\\
log $D_{\text{mix}}$ & $1.0_{-0.5}^{+0.8}$ & $1.5_{-0.8}^{+0.5}$ & $1.0_{-0.5}^{+0.8}$ & $1.0_{-0.5}^{+0.8}$ & $1.5_{-0.8}^{+0.5}$ & $1.0_{-0.5}^{+0.8}$\\
$X_{\text{c}}$ & $0.26_{-0.10}^{+0.08}$ & $0.24_{-0.10}^{+0.09}$ & $0.24_{-0.08}^{+0.10}$ & $0.26_{-0.10}^{+0.08}$ & $0.24_{-0.10}^{+0.09}$ & $0.24_{-0.08}^{+0.10}$\\
\hline
$R$ [$\text{R}_{\odot}$] & $2.0_{-0.2}^{+0.7}$ & $2.1_{-0.2}^{+0.8}$ & $2.1_{-0.2}^{+0.6}$ & $2.0_{-0.2}^{+0.7}$ & $2.1_{-0.2}^{+0.8}$ & $2.1_{-0.2}^{+0.6}$\\
$T_{\text{eff}}$ [K] & $7181_{-97}^{+130}$ & $7150_{-86}^{+113}$ & $7173_{-92}^{+123}$ & $7181_{-97}^{+130}$ & $7150_{-86}^{+113}$ & $7173_{-92}^{+123}$\\
log~g [dex] & $4.01_{-0.18}^{+0.06}$ & $3.97_{-0.20}^{+0.08}$ & $3.99_{-0.16}^{+0.07}$ & $4.01_{-0.18}^{+0.06}$ & $3.97_{-0.20}^{+0.08}$ & $3.99_{-0.16}^{+0.07}$\\
$M_{\text{cc}}$ [$M_{\odot}$] & $0.149_{-0.007}^{+0.008}$ & $0.148_{-0.006}^{+0.007}$ & $0.150_{-0.007}^{+0.007}$ & $0.149_{-0.007}^{+0.008}$ & $0.148_{-0.006}^{+0.007}$ & $0.150_{-0.007}^{+0.007}$\\
$\tau$ [Gyr] & $1.5_{-0.2}^{+0.3}$ & $1.5_{-0.2}^{+0.3}$ & $1.5_{-0.2}^{+0.4}$ & $1.5_{-0.2}^{+0.3}$ & $1.5_{-0.2}^{+0.3}$ & $1.5_{-0.2}^{+0.3}$\\
\hline
$\chi^{2}_{\text{red, min}}|\text{MD}_{\text{min}}$ & $1.13\cdot 10^{-13}$ & $6.79\cdot 10^{-15}$ & $7.58\cdot 10^{-7}$ & $3.73\cdot 10^{-17}$ & $3.18\cdot 10^{-15}$ & $7.11\cdot 10^{-8}$\\ 
 \hline
\end{tabular}
}
\tablefoot{Grid subset based on $6965\text{ K}\leq T_{\text{eff}}\leq 7520\text{ K}$ and $3.09\leq\text{log }g\leq4.13$.}
\end{center}
\label{tab: Pi0_as_spec}

\vspace{6pt}

\setlength{\tabcolsep}{6pt}
\renewcommand{\arraystretch}{1.3}
\caption[Same as Table \ref{tab: Pi0_as}, but for a grid subset based on the 3$\sigma$ interval of the dynamical $M$, $T_{\text{eff}}$ and log~g values.]{Same as Table \ref{tab: Pi0_as}, but for a grid subset based on the 3$\sigma$ interval of the dynamical $M$, $T_{\text{eff}}$ and log~g values (see footnote).}
\begin{center}
{\small
\begin{tabular}{ c|ccc|ccc } 
 \hline
 \hline
 & \multicolumn{3}{c|}{$\chi^{2}_{\text{red}}$} & \multicolumn{3}{c}{$\text{MD}$}\\
 \hline
  & $\bf{Y_{\rm obs}}=(\Pi_{0, \ \boldsymbol{\ell}=1})$ & $\bf{Y_{\rm obs}}=(\Pi_{0, \ \boldsymbol{\ell}=2})$  & $\bf{Y_{\rm obs}}=(\Pi_{0, \ \boldsymbol{\ell}=1}, \Pi_{0, \ \boldsymbol{\ell}=2})$ & $\bf{Y_{\rm obs}}=(\Pi_{0, \ \boldsymbol{\ell}=1})$ & $\bf{Y_{\rm obs}}=(\Pi_{0, \ \boldsymbol{\ell}=2})$  & $\bf{Y_{\rm obs}}=(\Pi_{0, \ \boldsymbol{\ell}=1}, \Pi_{0, \ \boldsymbol{\ell}=2})$\\ 
 \hline
$M$ [$\text{M}_{\odot}$] & $^{*}1.65_{-0.05}^{+0.05}$ & $^{*}1.65_{-0.05}^{+0.05}$ & $^{*}1.65_{-0.05}^{+0.05}$ & $^{*}1.65_{-0.05}^{+0.05}$  & $^{*}1.65_{-0.05}^{+0.05}$ & $^{*}1.65_{-0.05}^{+0.05}$\\
$f_{\text{ov}}$ & $^{*}0.005_{-0.005}^{+0.005}$ & $^{*}0.005_{-0.005}^{+0.005}$ & $^{*}0.005_{-0.005}^{+0.005}$ & $^{*}0.005_{-0.005}^{+0.005}$ & $^{*}0.005_{-0.005}^{+0.005}$ & $^{*}0.005_{-0.005}^{+0.005}$\\
log $D_{\text{mix}}$ & $0.5_{-0.3}^{+0.5}$ & $^{*}0.5_{-0.5}^{+0.5}$ & $^{*}0.5_{-0.5}^{+0.5}$ & $0.5_{-0.3}^{+0.5}$ & $^{*}0.5_{-0.5}^{+0.5}$ & $^{*}0.5_{-0.5}^{+0.5}$\\
$X_{\text{c}}$ & $0.22_{-0.02}^{+0.02}$ & $^{*}0.20_{-0.02}^{+0.02}$ & $^{*}0.20_{-0.02}^{+0.02}$ & $0.21_{-0.01}^{+0.01}$ & $^{*}0.20_{-0.02}^{+0.02}$ & $^{*}0.20_{-0.02}^{+0.02}$\\
\hline
$R$ [$\text{R}_{\odot}$] & $2.1_{-0.1}^{+0.1}$ & $2.2_{-0.1}^{+0.1}$ & $2.2_{-0.1}^{+0.1}$ & $2.2_{-0.1}^{+0.1}$ & $2.160_{-0.002}^{+0.002}$ & $2.160_{-0.002}^{+0.002}$\\
$T_{\text{eff}}$ [K] & $7462_{-196}^{+196}$ & $7435_{-196}^{+196}$ & $7435_{-196}^{+196}$ & $7458_{-196}^{+196}$ & $7435_{-196}^{+196}$ & $7435_{-196}^{+196}$\\
log~g [dex] & $3.99_{-0.04}^{+0.04}$ & $3.99_{-0.04}^{+0.04}$ & $3.99_{-0.04}^{+0.04}$ & $3.99_{-0.04}^{+0.04}$ & $3.99_{-0.04}^{+0.04}$ & $3.99_{-0.04}^{+0.04}$\\
$M_{\text{cc}}$ [$M_{\odot}$] & $0.150_{-0.001}^{+0.001}$ & $0.150_{-0.001}^{+0.001}$ & $0.150_{-0.001}^{+0.001}$ & $0.150_{-0.001}^{+0.001}$ & $0.150_{-0.001}^{+0.001}$ & $0.150_{-0.001}^{+0.001}$\\
$\tau$ [Gyr] & $1.2_{-0.2}^{+0.2}$ & $1.2_{-0.2}^{+0.2}$ & $1.2_{-0.2}^{+0.2}$ & $1.2_{-0.2}^{+0.2}$ & $1.2_{-0.2}^{+0.2}$ & $1.2_{-0.2}^{+0.2}$\\
\hline
$\chi^{2}_{\text{red, min}}|\text{MD}_{\text{min}}$ & $2.58\cdot 10^{-13}$ & $1.83\cdot 10^{-9}$ & $1.17\cdot 10^{-4}$ & $2.75\cdot 10^{-9}$ & $2.78\cdot 10^{-10}$ & $2.00\cdot 10^{-5}$\\ 
 \hline
\end{tabular}
}
\tablefoot{The $^{*}$ indicate the estimated parameters for which single-grid-step errors are assumed. Grid subset based on $1.63\text{ M}_{\odot}\leq M_{\text{ini}}\leq 1.70\text{ M}_{\odot}$, $7081\text{ K}\leq T_{\text{eff}}\leq 7590\text{ K}$ and $3.9826\leq\text{log }g\leq4.0021$.}
\end{center}
\label{tab: Pi0_as_dyn}
\end{table*}

It can be seen that there is generally little difference in the results regardless of whether a $\chi^{2}_{\text{red}}$ or the MD merit function is used, regardless of the grid setup or $\textbf{Y}$ configurations. This was an expected result, as the $\chi^{2}$ and MD merit functions converge for vectors of unit or near-unit length (due to the term $(\textbf{V}+\boldsymbol{\Lambda})^{-1}$ used in the computation of the MD, see Eq. \ref{eq: MD}). The small minima of the respective merit functions shown in Tables \ref{tab: Pi0_as} to \ref{tab: Pi0_as_dyn} are due to the near-perfect agreement (sub-second differences in the best cases) of the theoretical predictions with the perturbed observables in the Monte-Carlo framework, given that we fitted only one observable by the means of four free parameters. In addition, there is little difference in the results for the three different $\textbf{Y}$ configurations. This is also unsurprising, as it was already noted in Paper\,I that the calculated $\Pi_{0}$ values for each of these modes agree within 2$\sigma$. The fits using $\bf{Y_{\rm obs}}=(\Pi_{0, \ \boldsymbol{\ell}=1}, \Pi_{0, \ \boldsymbol{\ell}=2})$ are in agreement with those derived from evolutionary modelling (cf.\,Table \ref{tab: evol_params}). Overall, our results of purely asteroseismic modelling with $\textbf{Y}=(\bf{\Pi_{0}})$ provide weaker constraints on the external properties ($M$, $R$, $T_{\text{eff}}$ and log~g), superior constraints on the interior properties ($M_{\text{cc}}$, $f_{\text{ov}}$ and $D_{\text{mix}}$) and similar constraints on the evolutionary stage ($X_{\text{c}}$ and $\tau$) when compared to evolutionary modelling (cf. Tables \ref{tab: evol_params} and \ref{tab: Pi0_as}). This showcases the superiority of g-mode asteroseismic modelling over evolutionary modelling when it comes to constraining interior properties.  

The fits using the spectroscopic and dynamical grids (Tables \ref{tab: Pi0_as_spec} and \ref{tab: Pi0_as_dyn}) resulted in superior constraints on most parameters than using either purely asteroseismic or evolutionary modelling when using both the $\chi^{2}_{\text{red}}$ and MD merit functions. However, the best-fitting $\Pi_{0}$ values and errors are almost identical regardless of the grid subset used in the fit (Figure \ref{fig: Pi0_results}), demonstrating the value of asteroseismic observational constraints on parameter estimation due to its  degenerate nature. It can also be seen that the dynamical grid results cluster at the edge of the grid (see inset plot demarcated by a dotted box in Figure \ref{fig: Pi0_Kiel}), indicating that the minima of the respective merit functions are outside of the range of the dynamical grid. When compared to the full grid results, the spectroscopic and dynamical grid results also trended towards lower $f_{\text{ov}}$ and $D_{\text{mix}}$ values. 

As mentioned previously, the period-spacing patterns of KIC9850387 are relatively flat (i.e. do not display significant dips), implying either a high amount of mixing at the bottom of the radiative envelope or a young evolutionary stage (i.e. high $X_{\text{c}}$ values). In general, the envelope mixing levels in intermediate-mass g-mode pulsating stars are hard to infer, as investigated and argued by \cite{Mombarg2019}. Their sample study utilised principle component analysis to deduce an expectation value of $D_{\rm mix}=1\,$cm$^2$\,s$^{-1}$ for the 37 pulsators in their study, which is much lower than the wide range of values covered by B-type pulsators with much bigger convective cores \citep[][Table\,1]{Aerts2020}. Here, our results imply an evolved star with flat patterns revealing an above-average level of mixing for F-type g-mode pulsators ($D_{\rm mix}\sim 3\,$cm$^2$\,s$^{-1}$). In fact, given the flat period-spacing patterns, this value should be seen as a lower limit since higher envelope mixing will not change the already flat spacing patterns and does not change the global parameters of the stars \citep{Mombarg2019}. 

The minima of the merit functions listed in Tables\,\ref{tab: Pi0_as},
\ref{tab: Pi0_as_spec}, and \ref{tab: Pi0_as_dyn} reveal that using just one observable to estimate the four free parameters ($M$, $X_{\text{c}}$, $f_{\rm ov}$ and $\log D_{\rm mix}$) is not very constraining as there are too many degrees of freedom. It is therefore of interest to fit the individual $\Delta P_{\ell}$ values of the observed dipole and quadrupole modes.

\subsection{$\Delta P_{\ell}$ modelling results}
\label{subsec: PS}

\begin{figure*}[ht!]
\includegraphics[width=\hsize]{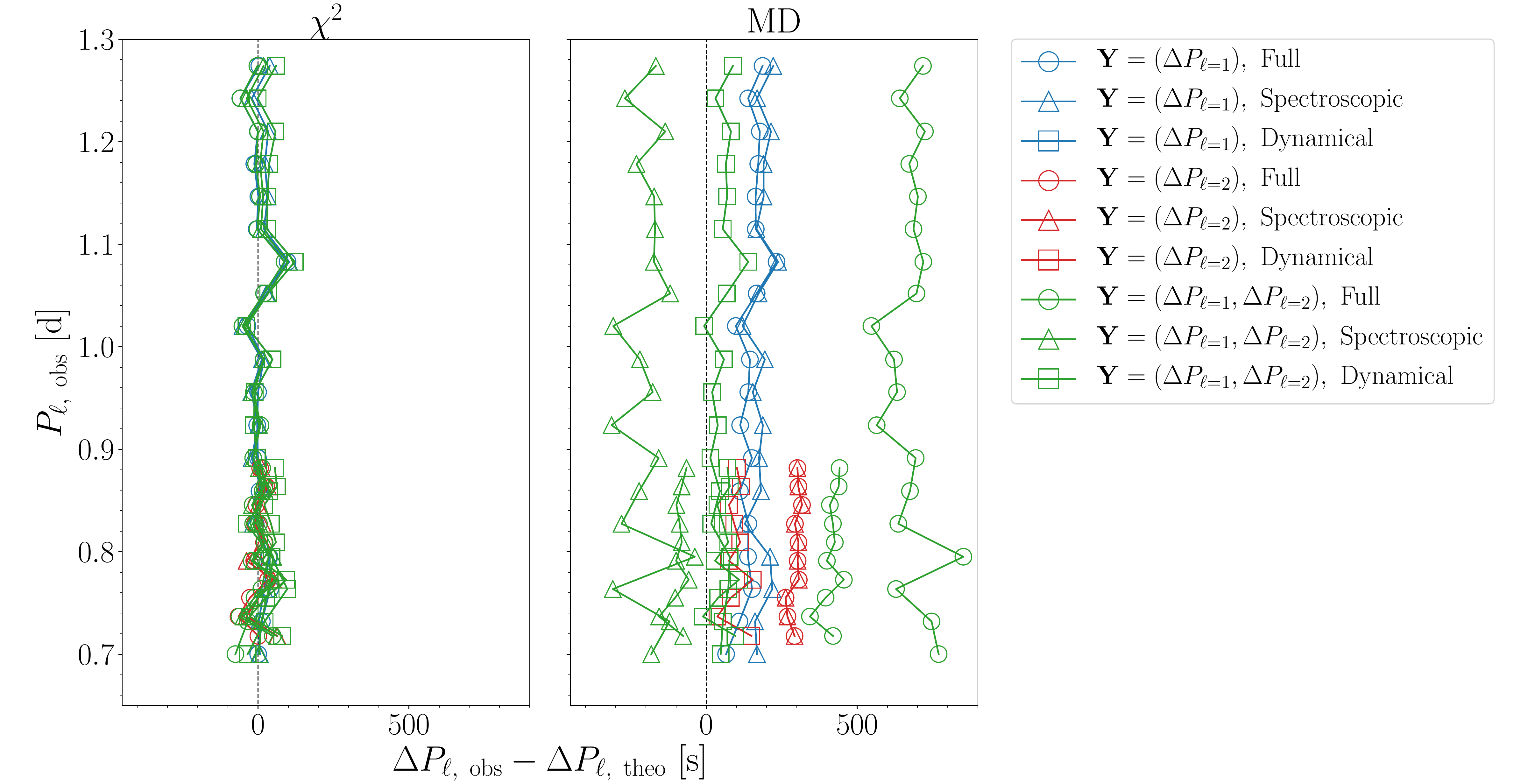}
\caption[The best-fitting $\Delta P_{\ell}$ values of our $\Delta P_{\ell}$-based asteroseismic modelling.]{Same as Figure \ref{fig: Pi0_results}, but representing the results of $\Delta P_{\ell}$-based asteroseismic modelling.}
\label{fig: PS_results}

\vspace{24pt}

\includegraphics[width=\hsize]{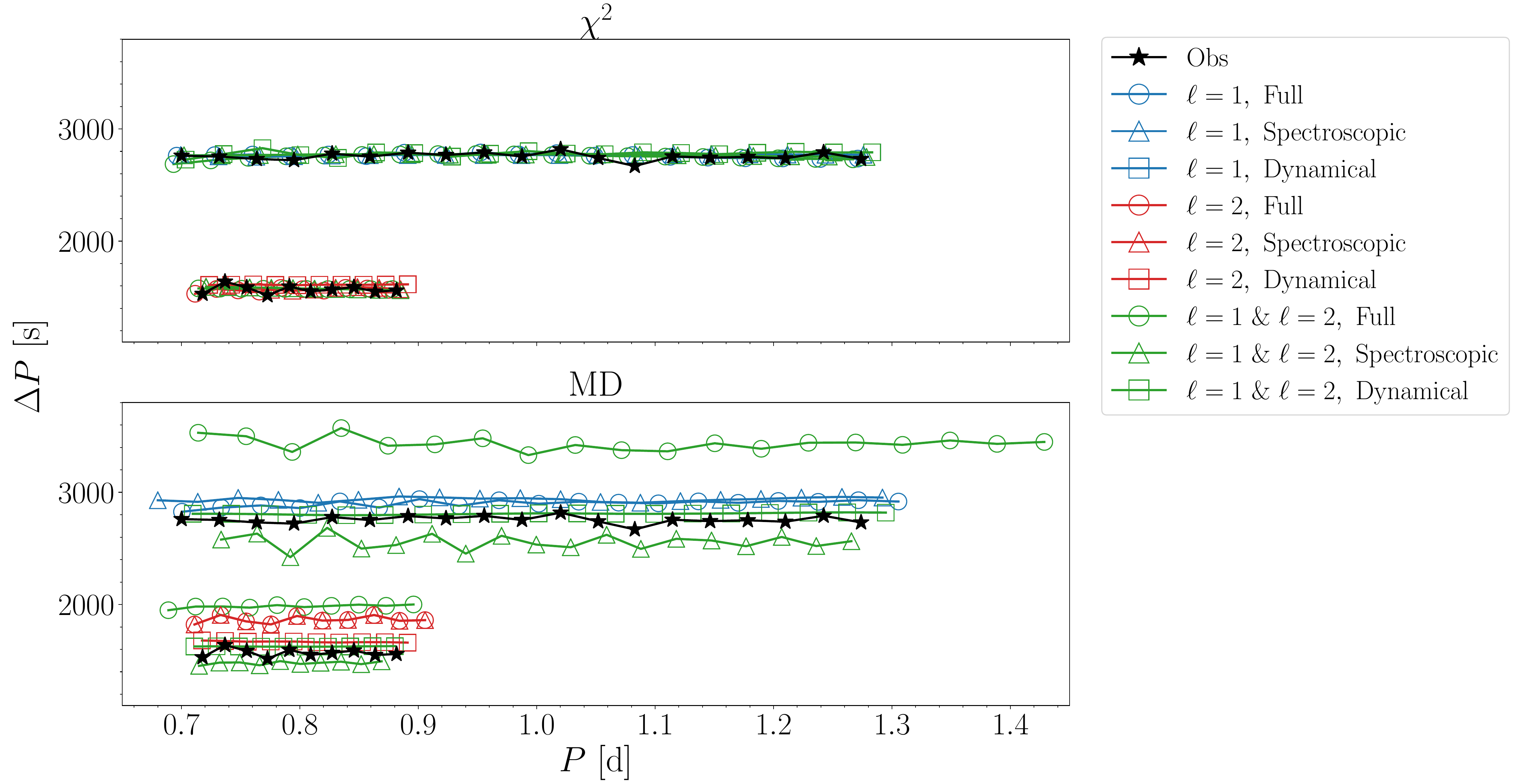}
\caption[Best-fitting $\ell=1$ and $\ell=2$ period-spacing patterns using our $\Delta P_{\ell}$-based asteroseismic modelling.]{Best-fitting $\ell=1$ and $\ell=2$ period-spacing patterns using our $\Delta P_{\ell}$-based asteroseismic modelling. The top panel shows the results of using a $\chi^{2}_{\text{red}}$ merit function and the bottom panel shows the results of using an MD merit function.}
\label{fig: PS_bestfit}
\end{figure*}

\begin{figure*}[!ht]
\includegraphics[width=\hsize]{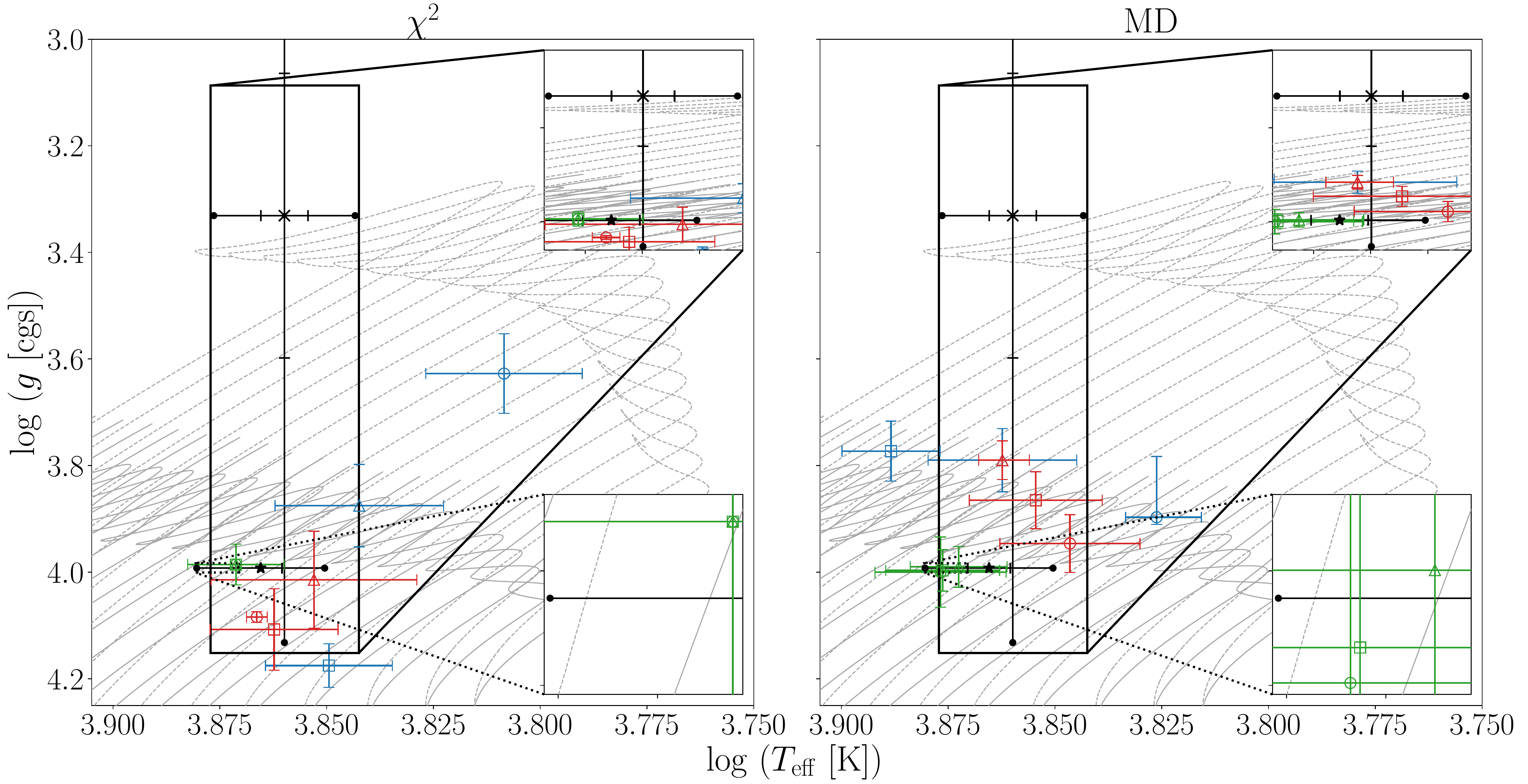}
\caption[Positions of the best-fitting models from $\Delta P_{\ell}$-based asteroseismic modelling on a Kiel diagram.]{Same as Figure \ref{fig: Pi0_Kiel}, but representing the results of $\Delta P_{\ell}$-based asteroseismic modelling.}
\label{fig: PS_Kiel}
\end{figure*}

We performed our $\Delta P_{\ell}$ modelling in an identical way to our $\Pi_{0}$ modelling, based on the three grid subsets (full, spectroscopic and dynamical), using three parameter setups (${\textbf{Y}=(\Delta P_{\ell=1, \ i})}$, ${\textbf{Y}=(\Delta P_{\ell=2, \ j})}$ and ${\textbf{Y}=(\Delta P_{\ell=1, \ i}, \Delta P_{\ell=2, \ j})}$, where $i$ and $j$ are the indices of the $\Delta P_{\ell}$ values of the corresponding period-spacing patterns), and computed the same two merit functions ($\chi^{2}_{\text{red}}$ and MD). These results are listed in Table \ref{tab: PS_as} (full grid), Table \ref{tab: PS_as_spec} (spectroscopic grid) and Table \ref{tab: PS_as_dyn} (dynamical grid). Comparisons of the $\chi^{2}_{\text{red}}$ and the MD results in terms of the best-fitting $\Delta P_{\ell}$ values, period-spacing patterns and positions on Kiel diagrams are shown in Figures \ref{fig: PS_results}, \ref{fig: PS_bestfit} and \ref{fig: PS_Kiel} respectively. 

During our best solution and precision estimation, we encountered
a number of solutions whose precision could not be assessed, because 
the same grid point was selected in every Monte-Carlo iteration (as described in Section \ref{sec: modelling}). These so-called 'single-grid-point' solutions occur more frequently for the $\chi^{2}_{\text{red}}$ (7/9 solutions over the three grids) than for the MD framework (5/9 solutions over the three grids). This is due to the MD incorporating the uncertainties in the theoretical predictions, offering a broader range of acceptable solutions to match the perturbations of the ${\textbf{Y}_{\text{obs}}}$ vector. In these cases, we once again assigned single-grid-point errors to the estimated parameters and propagated them to the inferred parameters. Similar to our purely asteroseismic $\Pi_{0}$-based results, $\Delta P_{\ell}$-based modelling resulted in weaker constraints on the external properties ($M$, $R$, $T_{\text{eff}}$ and log~g) and superior constraints on the interior properties ($M_{\text{cc}}$, $f_{\text{ov}}$ and $D_{\text{mix}}$) when compared to evolutionary modelling. 

The effect of the choice of merit function is demonstrated by Figure \ref{fig: PS_results}. The MD solutions offer a larger variety of appropriate solutions according the various MD minima, because this metric allows for theoretical uncertainty in the modelling. Since the $\chi^{2}_{\text{red}}$ is a specific solution of the regression based on the assumption that there are no theoretical imperfections, while ignoring correlations among the observables and parameters, its best solutions always occur in the MD solution space as well, but often at much higher MD values. This is graphically illustrated in Figure \ref{fig: PS_results}, where the broad coverage of the best allowed period spacing values selected by the MD is due to the allowance of uncertainty in the theoretical models via the variance covered by the entire grid. For example, the MD value of the model from the dynamical grid (Table \ref{tab: PS_as_dyn}) with the lowest $\chi^{2}_{\text{red}}$ value ($\chi^{2}_{\text{red}}=597$) is 2674, which is larger than the lowest MD value of 2378. However, the overall best $\chi^{2}_{\text{red}}$ and MD solutions from the dynamical grid, demanding compliance with both the $\Delta P_{\ell}$ values of the dipole and quadrupole modes, binary and the spectroscopy, are very similar. The MD methodology is sensitive to the model-independent constraints imposed upon the problem set, as it allows for the theory of the used grid to be imperfect at the level captured in the variance-covariance matrix induced by the free parameter ranges. Nevertheless, both metrics end up with an almost equal solution when demanding compliance with all observational constraints at the $3\sigma$ level. The overall best solution combining all of the asteroseismic and dynamical constraints is indicated by the MD solution in boldface in Table \ref{tab: PS_as_dyn}.

Similar to our $\Pi_{0}$-based results, our dynamical grid results imply above-average levels of envelope mixing, and in the ${\textbf{Y}=(\Delta P_{\ell=1, \ i})}$ and ${\textbf{Y}=(\Delta P_{\ell=1, \ i}, \Delta P_{\ell=2, \ j})}$ cases we find high values of $D_{\rm mix}\simeq 25\,$cm$^2$\,s$^{-1}$ compared to those derived for single $\gamma\,$Dor pulsators by \citep{Mombarg2019}. While this may indicate extra tidal mixing, casting doubt on the use of eclipsing binaries as testbeds of stellar structural and asteroseismic theory of single stars, several slowly-rotating single stars also display near-flat period-spacing patterns \citep{Li2019a} as was found for KIC9850387. Hence we posit that the estimated level of envelope mixing could still be due to intrinsic non-tidal element transport mechanisms.

\begin{table*}[!htp]
\setlength{\tabcolsep}{6pt}
\renewcommand{\arraystretch}{1.3}
\caption[Results of asteroseismic modelling with ${\textbf{Y}=(\Delta P_{\ell})}$ using the full grid of pulsational models.]{Same as Table \ref{tab: Pi0_as_dyn}, but representing the results of our asteroseismic modelling with ${\textbf{Y}=(\Delta P_{\ell})}$ using the full grid of pulsational models.}
\begin{center}
{\small
\begin{tabular}{ c|ccc|ccc } 
 \hline
 \hline
 & \multicolumn{3}{c|}{$\chi^{2}_{\text{red}}$} & \multicolumn{3}{c}{$\text{MD}$}\\
 \hline
  & $\textbf{Y}=(\Delta P_{\ell=1, \ i})$ & $\textbf{Y}=(\Delta P_{\ell=2, \ j})$  & $\textbf{Y}=(\Delta P_{\ell=1, \ i}, \Delta P_{\ell=2, \ j})$ & $\textbf{Y}=(\Delta P_{\ell=1, \ i})$ & $\textbf{Y}=(\Delta P_{\ell=2, \ j})$  & $\textbf{Y}=(\Delta P_{\ell=1, \ i}, \Delta P_{\ell=2, \ j})$ \\ 
 \hline
$M$ [$\text{M}_{\odot}$] & $^{*}1.65_{-0.05}^{+0.05}$ & $^{*}1.60_{-0.05}^{+0.05}$ & $^{*}1.35_{-0.05}^{+0.05}$ & $1.50_{-0.05}^{+0.13}$  & $^{*}1.70_{-0.05}^{+0.05}$ & $^{*}1.85_{-0.05}^{+0.05}$\\
$f_{\text{ov}}$ & $^{*}0.030_{-0.005}^{+0.005}$ & $^{*}0.015_{-0.005}^{+0.005}$ & $^{*}0.035_{-0.005}^{+0.005}$ & $0.015_{-0.005}^{+0.005}$ & $^{*}0.040_{-0.005}^{+0.005}$ & $^{*}0.040_{-0.005}^{+0.005}$\\
log $D_{\text{mix}}$ & $^{*}0.5_{-0.5}^{+0.5}$ & $^{*}0.0^{+0.5}$ & $^{*}1.5_{-0.5}^{+0.5}$ & $0.1_{-0.1}^{+0.1}$ & $^{*}0.0^{+0.5}$ & $^{*}0.5_{-0.5}^{+0.5}$\\
$X_{\text{c}}$ & $^{*}0.10_{-0.02}^{+0.02}$ & $^{*}0.16_{-0.02}^{+0.02}$ & $^{*}0.48_{-0.02}^{+0.02}$ & $0.10_{-0.02}^{+0.02}$ & $^{*}0.24_{-0.02}^{+0.02}$ & $^{*}0.24_{-0.02}^{+0.02}$\\
\hline
$R$ [$\text{R}_{\odot}$] & $3.3_{-0.3}^{+0.3}$ & $2.4_{-0.3}^{+0.3}$ & $1.6_{-0.1}^{+0.1}$ & $2.28_{-0.05}^{+0.42}$ & $2.7_{-0.2}^{+0.2}$ & $2.9_{-0.2}^{+0.2}$\\
$T_{\text{eff}}$ [K] & $6434_{-277}^{+277}$ & $6956_{-323}^{+323}$ & $7071_{-246}^{+246}$ & $6702_{-164}^{+113}$ & $7283_{-298}^{+298}$ & $7734_{-207}^{+207}$\\
log~g [dex] & $3.63_{-0.07}^{+0.07}$ & $3.88_{-0.08}^{+0.08}$ & $4.18_{-0.04}^{+0.04}$ & $3.90_{-0.11}^{+0.01}$ & $3.79_{-0.06}^{+0.06}$ & $3.77_{-0.06}^{+0.06}$\\
$M_{\text{cc}}$ [$M_{\odot}$] & $0.142_{-0.02}^{+0.02}$ & $0.14_{-0.1}^{+0.1}$ & $0.1_{-0.4}^{+0.4}$ & $0.118_{-0.005}^{+0.020}$ & $0.19_{-0.02}^{+0.02}$ & $0.22_{-0.02}^{+0.02}$\\
$\tau$ [Gyr] & $1.7_{-0.2}^{+0.2}$ & $1.5_{-0.3}^{+0.3}$ & $1.7_{-0.5}^{+0.5}$ & $1.7_{-0.1}^{+0.3}$ & $1.5_{-0.2}^{+0.2}$ & $1.22_{-0.08}^{+0.08}$\\
 \hline
$\chi^{2}_{\text{red, min}}|\text{MD}_{\text{min}}$ & $187$ & $292$ & $333$ & $19$ & $20$ & $153$\\ 
 \hline
\end{tabular}
}
\end{center}
\label{tab: PS_as}

\setlength{\tabcolsep}{6pt}
\renewcommand{\arraystretch}{1.3}
\caption[Same as Table \ref{tab: PS_as}, but for a grid subset based on the 3$\sigma$ interval of the `pseudo-single-star' spectroscopic $T_{\text{eff}}$ and log~g values.]{Same as Table \ref{tab: PS_as}, but for a grid subset based on the 3$\sigma$ interval of the `pseudo-single-star' spectroscopic $T_{\text{eff}}$ and log~g values (see footnote).}
\begin{center}
{\small
\begin{tabular}{ c|ccc|ccc }
 \hline
 \hline
 & \multicolumn{3}{c|}{$\chi^{2}_{\text{red}}$} & \multicolumn{3}{c}{$\text{MD}$}\\
 \hline
  & $\textbf{Y}=(\Delta P_{\ell=1, \ i})$ & $\textbf{Y}=(\Delta P_{\ell=2, \ j})$  & $\textbf{Y}=(\Delta P_{\ell=1, \ i}, \Delta P_{\ell=2, \ j})$ & $\textbf{Y}=(\Delta P_{\ell=1, \ i})$ & $\textbf{Y}=(\Delta P_{\ell=2, \ j})$  & $\textbf{Y}=(\Delta P_{\ell=1, \ i}, \Delta P_{\ell=2, \ j})$ \\ 
  \hline
$M$ [$\text{M}_{\odot}$] & $1.51_{-0.01}^{+0.01}$ & $1.53_{-0.08}^{+0.08}$ & $^{*}1.45_{-0.05}^{+0.05}$ & $^{*}1.50_{-0.05}^{+0.05}$  & $1.71_{-0.01}^{+0.01}$ & $^{*}1.70_{-0.05}^{+0.05}$ \\
$f_{\text{ov}}$ & $0.009_{-0.001}^{+0.001}$ & $0.014_{-0.001}^{+0.001}$ & $^{*}0.015_{-0.005}^{+0.005}$ & $^{*}0.035_{-0.005}^{+0.005}$ & $0.038_{-0.003}^{+0.003}$ & $^{*}0.005_{-0.005}^{+0.005}$\\
log $D_{\text{mix}}$ & $1.4_{-0.1}^{+0.1}$ & $^{*}0.0^{+0.5}$ & $^{*}2.0_{-0.5}^{+0.5}$ & $^{*}0.5_{-0.5}^{+0.5}$ & $^{*}0.0^{+0.5}$ & $^{*}0.0^{+0.5}$\\
$X_{\text{c}}$ & $0.33_{-0.02}^{+0.02}$ & $0.27_{-0.06}^{+0.06}$ & $^{*}0.40_{-0.02}^{+0.02}$ & $^{*}0.30_{-0.02}^{+0.2}$ & $0.23_{-0.01}^{+0.01}$ & $^{*}0.08_{-0.02}^{+0.02}$\\
\hline
$R$ [$\text{R}_{\odot}$] & $1.85_{-0.01}^{+0.01}$ & $2.0_{-0.3}^{+0.3}$ & $1.8_{-0.2}^{+0.2}$ & $^{*}2.2_{-0.2}^{+0.2}$ & $2.7_{-0.1}^{+0.1}$ & $2.5_{-0.2}^{+0.2}$\\
$T_{\text{eff}}$ [K] & $7351_{-28}^{+28}$ & $7127_{-406}^{+406}$ & $7283_{-256}^{+256}$ & $7023_{-271}^{+271}$ & $7283_{-106}^{+93}$ & $7153_{-261}^{+261}$\\
log~g [dex] & $4.084_{-0.001}^{+0.001}$ & $4.01_{-0.09}^{+0.09}$ & $4.11_{-0.08}^{+0.08}$ & $3.94_{-0.05}^{+0.05}$ & $3.79_{-0.04}^{+0.04}$ & $3.86_{-0.05}^{+0.05}$\\
$M_{\text{cc}}$ [$M_{\odot}$] & $0.1519_{-0.0007}^{+0.0007}$ & $0.14_{-0.02}^{+0.02}$ & $0.156_{-0.03}^{+0.03}$ & $0.16_{-0.01}^{+0.01}$ & $0.188_{-0.006}^{+0.006}$ & $0.13_{-0.01}^{+0.01}$\\
$\tau$ [Gyr] & $1.38_{-0.05}^{+0.03}$ & $1.6_{-0.4}^{+0.3}$ & $1.2_{-0.6}^{+0.6}$ & $1.9_{-0.3}^{+0.3}$ & $1.50_{-0.09}^{+0.05}$ & $1.35_{-0.2}^{+0.2}$\\
 \hline
$\chi^{2}_{\text{red, min}}|\text{MD}_{\text{min}}$ & $243$ & $399$ & $342$ & $11$ & $17$ & $93$\\ 
 \hline
\end{tabular}
}
\tablefoot{Grid subset based on $6965\text{ K}\leq T_{\text{eff}}\leq 7520\text{ K}$ and $3.09\leq\text{log }g\leq4.13$}
\end{center}
\label{tab: PS_as_spec}

\setlength{\tabcolsep}{6pt}
\renewcommand{\arraystretch}{1.3}
\caption[Same as Table \ref{tab: PS_as}, but for a grid subset based on the 3$\sigma$ interval of the dynamical $M$, $T_{\text{eff}}$ and log~g values.]{Same as Table \ref{tab: PS_as}, but for a grid subset based on the 3$\sigma$ interval of the dynamical $M$, $T_{\text{eff}}$ and log~g values (see footnote).}
\begin{center}
{\small
\begin{tabular}{ c|ccc|ccc } 
 \hline
 \hline
 & \multicolumn{3}{c|}{$\chi^{2}_{\text{red}}$} & \multicolumn{3}{c}{$\text{MD}$}\\
 \hline
  & $\textbf{Y}=(\Delta P_{\ell=1, \ i})$ & $\textbf{Y}=(\Delta P_{\ell=2, \ j})$  & $\textbf{Y}=(\Delta P_{\ell=1, \ i}, \Delta P_{\ell=2, \ j})$ & $\textbf{Y}=(\Delta P_{\ell=1, \ i})$ & $\textbf{Y}=(\Delta P_{\ell=2, \ j})$  & $\textbf{Y}=(\Delta P_{\ell=1, \ i}, \Delta P_{\ell=2, \ j})$ \\ 
 \hline
$M$ [$\text{M}_{\odot}$] & $^{*}1.65_{-0.05}^{+0.05}$ & $^{*}1.65_{-0.05}^{+0.05}$ & $^{*}1.65_{-0.05}^{+0.05}$ & $^{*}1.65_{-0.05}^{+0.05}$  & $^{*}1.65_{-0.05}^{+0.05}$ & $\bf{^{*}1.65_{-0.05}^{+0.05}}$\\
$f_{\text{ov}}$ & $^{*}0.005_{-0.005}^{+0.005}$ & $^{*}0.005_{-0.005}^{+0.005}$ & $^{*}0.005_{-0.005}^{+0.005}$ & $^{*}0.005_{-0.005}^{+0.005}$ & $^{*}0.006_{-0.005}^{+0.005}$ & $^{*}\bf{0.006_{-0.005}^{+0.005}}$\\
log $D_{\text{mix}}$ & $^{*}0.5_{-0.5}^{+0.5}$ & $^{*}0.5_{-0.5}^{+0.5}$ & $^{*}0.5_{-0.5}^{+0.5}$ & $^{*}1.5_{-0.5}^{+0.5}$ & $0.1_{-0.1}^{+0.1}$ & $\bf{1.4_{-0.1}^{+0.1}}$\\
$X_{\text{c}}$ & $^{*}0.20_{-0.02}^{+0.02}$ & $^{*}0.20_{-0.02}^{+0.02}$ & $^{*}0.20_{-0.02}^{+0.02}$ & $^{*}0.24_{-0.02}^{+0.02}$ & $0.21_{-0.01}^{+0.01}$ & $\bf{^{*}0.24_{-0.02}^{+0.02}}$\\
\hline
$R$ [$\text{R}_{\odot}$] & $2.2_{-0.1}^{+0.1}$ & $2.2_{-0.1}^{+0.1}$ & $2.2_{-0.1}^{+0.1}$ & $2.1_{-0.2}^{+0.2}$ & $2.2_{-0.1}^{+0.1}$ & $\bf{2.1_{-0.1}^{+0.1}}$\\
$T_{\text{eff}}$ [K] & $7434_{-196}^{+196}$ & $7434_{-196}^{+196}$ & $7434_{-196}^{+196}$ & $7530_{-271}^{+271}$ & $7457_{-187}^{+187}$ & $\bf{7521_{-234}^{+234}}$\\
log~g [dex] & $3.99_{-0.04}^{+0.04}$ & $3.99_{-0.04}^{+0.04}$ & $3.99_{-0.04}^{+0.04}$ & $4.00_{-0.07}^{+0.07}$ & $3.99_{-0.03}^{+0.03}$ & $\bf{4.00_{-0.04}^{+0.04}}$\\
$M_{\text{cc}}$ [$M_{\odot}$] & $0.15_{-0.01}^{+0.01}$ & $0.15_{-0.01}^{+0.01}$ & $0.15_{-0.01}^{+0.01}$ & $0.16_{-0.02}^{+0.02}$ & $0.15_{-0.01}^{+0.01}$ & $\bf{0.16_{-0.01}^{+0.01}}$\\
$\tau$ [Gyr] & $1.2_{-0.2}^{+0.2}$ & $1.2_{-0.2}^{+0.2}$ & $1.2_{-0.2}^{+0.2}$ & $1.2_{-0.2}^{+0.2}$ & $1.2_{-0.1}^{+0.1}$ & $\bf{1.2_{-0.1}^{+0.1}}$\\
 \hline
$\chi^{2}_{\text{red, min}}|\text{MD}_{\text{min}}$ & $597$ & $1207$ & $927$ & $2403$ & $2937$ & $7555$\\ 
 \hline
\end{tabular}
}
\tablefoot{The overall best solution of our asterosesimic analyses is indicated in bold. Grid subset based on $1.63\text{ M}_{\odot}\leq M_{\text{ini}}\leq 1.70\text{ M}_{\odot}$, $7081\text{ K}\leq T_{\text{eff}}\leq 7590\text{ K}$ and $3.9826\leq\text{log }g\leq4.0021$.}
\end{center}
\label{tab: PS_as_dyn}
\end{table*}

\section{Investigating high-frequency modes}
\label{sec: HF_modes}

\begin{figure}[t]
\includegraphics[width=\hsize]{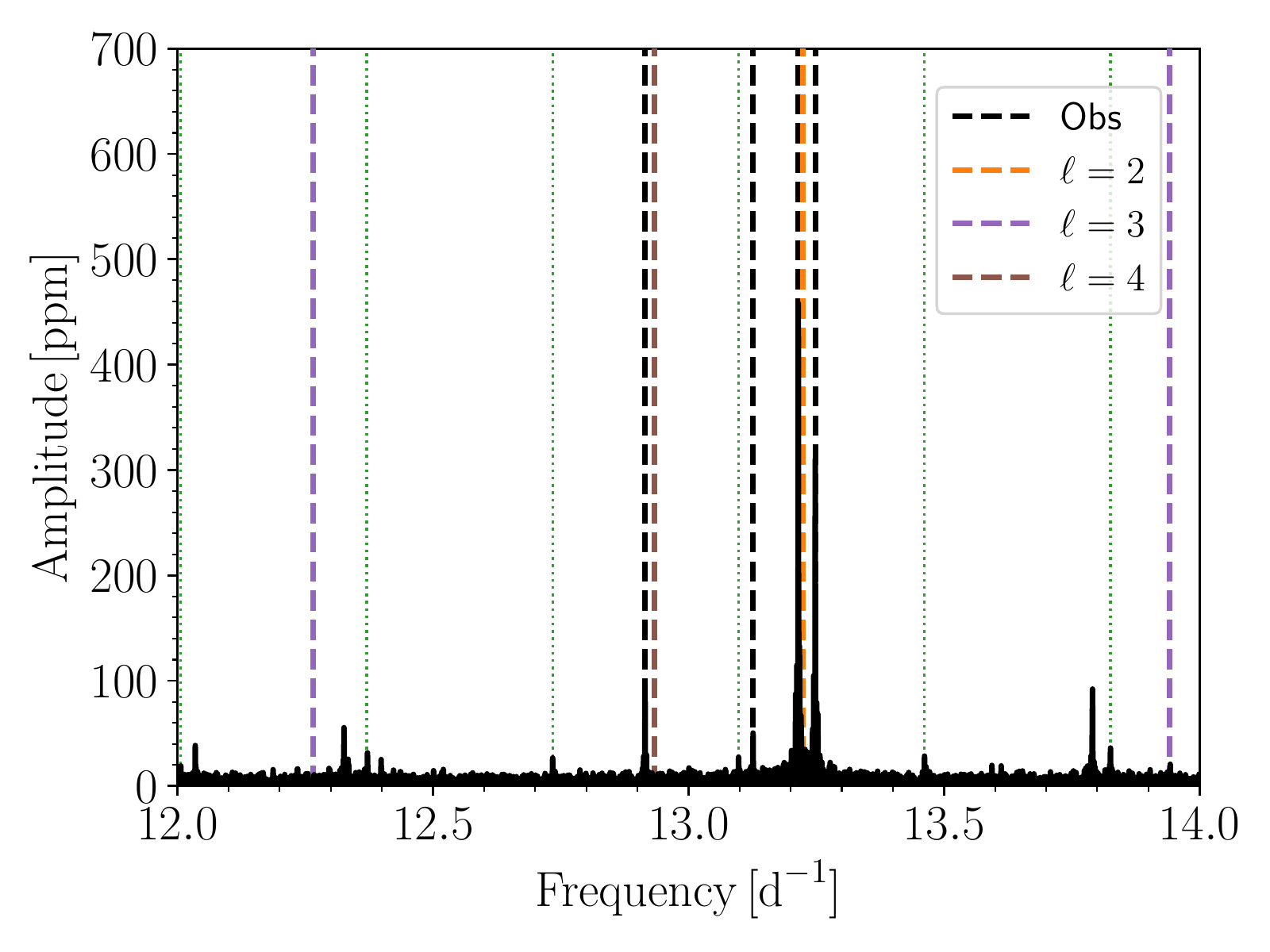}
\caption[Pulsational model with the closest match between the observed and theoretical modes in the high-frequency regime of KIC9850387.]{Pulsational model with the closest frequency match between the observed and theoretical modes in the high-frequency regime of KIC9850387. The vertical dotted green lines represent the orbital harmonics.}
\label{fig: Highfreqs_freqmatch}
\end{figure}

As noted in Paper\,I, a handful of independent high-frequency modes were observed in the frequency spectrum of KIC9850387. It was also stated that no frequency splitting or characteristic spacing was observed. Having obtained theoretical models that explain our observed $\Delta P_{\ell}$ values, we test that claim by computing theoretical frequencies of the $\ell=0$ to $\ell=4$ modes in the 12.8~d$^{-1}$ to 13.4~d$^{-1}$ frequency range for the stellar structural models within the overall best MD solution indicated in bold in Table \ref{tab: PS_as_dyn}. It was found that only two or three frequencies were obtained per structural model within the considered range, and the modes associated with these frequencies have varying $\ell$ values, with modes of certain $\ell$ values not occurring in the observed range. For example, the pulsational model with the closest match between the theoretical and observational frequencies, with parameters $M=1.70$~M$_{\odot}$, $X_{\text{c}}=0.22$, $f_{\rm ov}=0.01$ and $\log D_{\rm mix}=1.5$, consists an $\ell=2$ mixed mode\footnote{A mixed mode is a type of pulsational mode with p-mode character in the thin convective envelope and g-mode character in the radiative interior. See e.g. \cite{Aerts2010} for more information.} (i.e. $n_{\text{p}}=1$, $n_{\text{g}}=3$) at a frequency of 13.22~d$^{-1}$ and a low-order $\ell=4$ mode (i.e. $n_{\text{g}}=5$) at a frequency of 12.93~d$^{-1}$ (see Figure \ref{fig: Highfreqs_freqmatch}, where we show all modes occurring in the 12.0~d$^{-1}$ to 14.0~d$^{-1}$ frequency range).
 
It can be seen that the theoretical modes are offset from the observed modes. The same phenomenon was also observed for the F-type binary pulsator KIC10080943 \citep{Schmid2016}. Following these authors, we also posit that this offset is due to a so-called surface effect (see \citealt{Ball2017b} for a recent review) that is commonplace in solar-like stars due to uncertainties in the physical descriptions of the near-surface convection and non-adiabatic effects in the outer envelope. Indeed,  while intermediate-mass stars such as the primary component of KIC9850387 have thin convective envelopes, it has been shown that time-dependent convection theory is required to model the oscillations properly (see \citealt{Dupret2005}). The \textsc{mesa} and \textsc{gyre} codes do not include a treatment of time-dependent convection. Hence, frequency shifts between the theoretical predictions and the observed modes are expected given that the best models were fitted to the g~modes. Their mode energy is determined by the physics in the deep adiabatic interior of the star, where the approximation of time-independent convection is appropriate. 

The pulsational model with the closest match between the theoretical and observational frequencies can only potentially explain two out of the four observational frequencies. In order to explain more of the observational behaviour, we hypothesise that the pair of observed high-amplitude modes at 13.21~d$^{-1}$ and 13.25~d$^{-1}$ displayed in Figure \ref{fig: Highfreqs_freqmatch} is part of a rotationally split multiplet. However, the closest theoretical frequency is the $\ell=2$ mixed mode at 13.22~d$^{-1}$, and rotational splitting of such a mode would result in a quintuplet.
Therefore, it is more likely that the doublet is a rotationally-split prograde-retrograde ($m=1$ and $m=-1$) doublet with a missing zonal ($m=0$) mode. This phenomenon is a result of cancellation effects \citep{Aerts2010} due to the near 90$^{\circ}$ inclination of the system, and has been observed previously in hybrid p- and g-mode pulsators in the mass range of KIC9850387, most notably in \cite{Kurtz2014} where multiple well-resolved complete multiplets were reported.

\begin{figure}[t]
\includegraphics[width=\hsize]{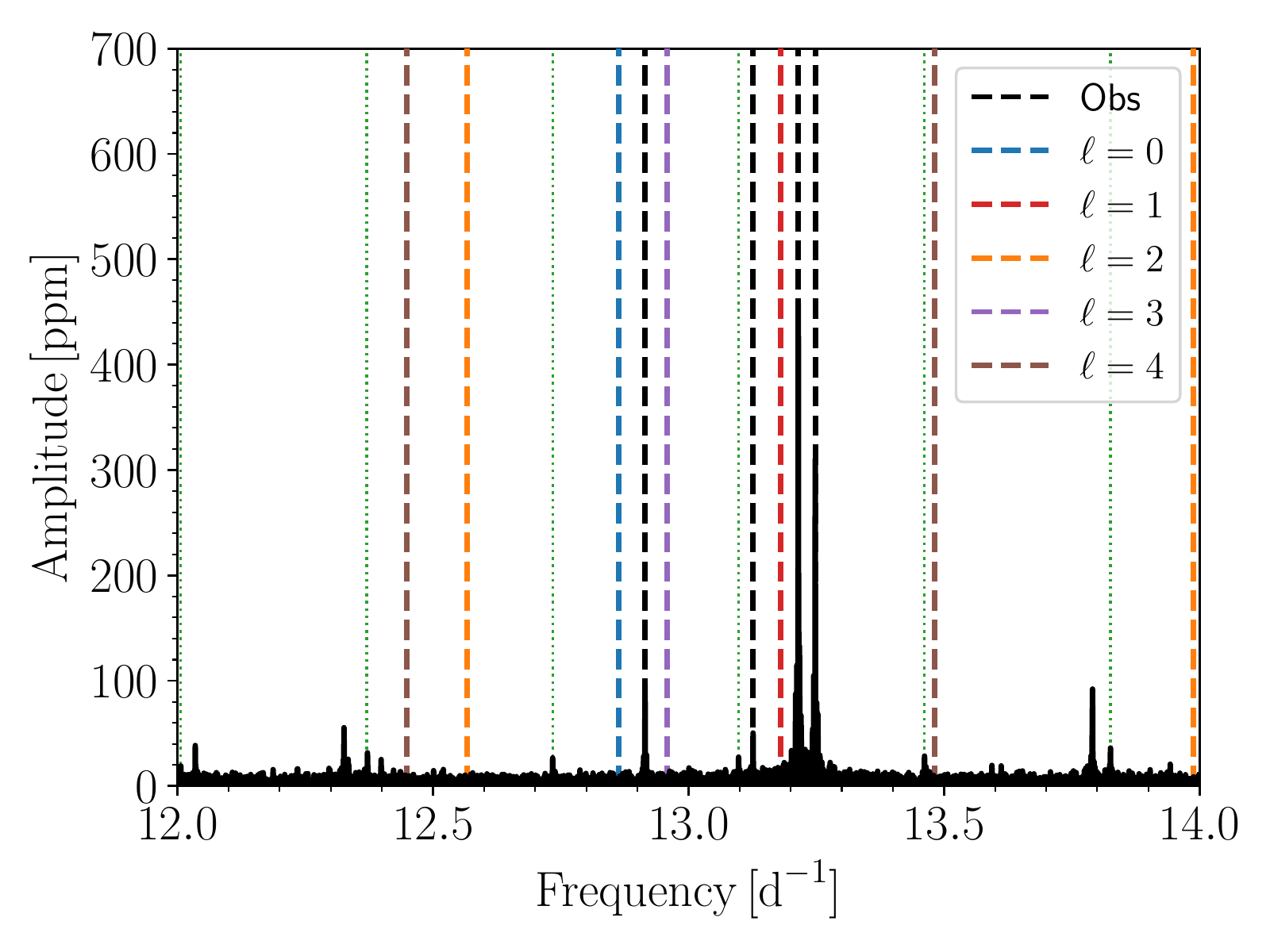}
\caption[Pulsational model with the closest frequency match between the theoretical $\ell=1$ mode and the midpoint of the high-amplitude doublet at 13.21~d$^{-1}$ and 13.25~d$^{-1}$.]{Pulsational model with the closest frequency match between the theoretical $\ell=1$ mode and the midpoint of the high-amplitude doublet at 13.21~d$^{-1}$ and 13.25~d$^{-1}$. The vertical dotted green lines represent the orbital harmonics.}
\label{fig: Highfreqs_doublet}
\end{figure}

As such, we further restricted our pulsational models to those that included $\ell=1$ modes within the 12.8~d$^{-1}$ to 13.4~d$^{-1}$ frequency range. We then determined the model with the smallest difference in frequency between the $\ell=1$ mode and the midpoint of the observed frequency doublet (13.23~d$^{-1}$). This best model has parameters $M=1.65$~M$_{\odot}$, $X_{\text{c}}=0.24$, $f_{\rm ov}=0.005$ and $\log D_{\rm mix}=1.5$, and three modes were obtained within this range: The fundamental radial mode (i.e. $n_{\text{p}}=1$, $\ell=0$) at a frequency of 12.86~d$^{-1}$, a low-order octupole g mode (i.e. $n_{\text{g}}=4$, $\ell=3$) at a frequency of 12.96~d$^{-1}$ and the ${\text{g}_1}$ mode (i.e. $n_{\text{g}}=1$, $\ell=1$) at a frequency of 13.18~d$^{-1}$. These theoretical mode predictions, along with the observed modes, are displayed in Figure \ref{fig: Highfreqs_doublet}. It should be noted that modes with $\ell>2$ are generally not observed in space-based photometry due to geometric cancellation effects \citep{Aerts2010}, although there have been claims to the contrary (see e.g. \citealt{Baran2013}). As such, we hypothesise that the observed mode at 12.92~d$^{-1}$ is the fundamental radial mode. The fundamental radial mode is offset from the nearest observed mode by $\sim$0.05~d$^{-1}$ while the dipole g$_1$ mode is offset from the nearest observed mode by $\sim$0.03~d$^{-1}$. The values of these frequency offsets are smaller than those observed for KIC10080943 \citep{Schmid2016}, who were unable to identify the degree of the p~modes for KIC10080943.

\begin{figure}[t]
\includegraphics[width=\hsize]{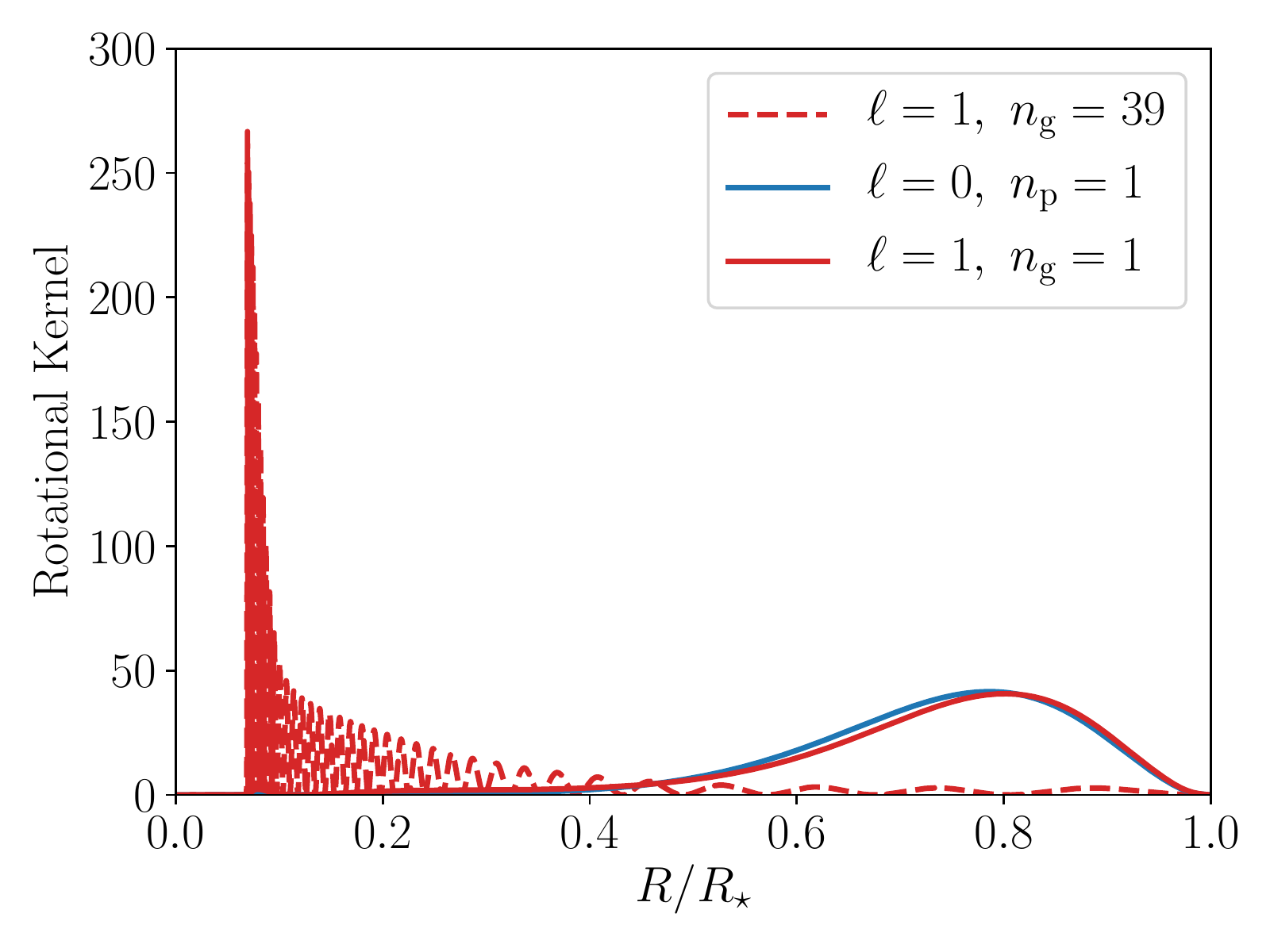}
\caption[Rotational kernels of the high and low order modes of the best-fitting pulsational model as a function of the fractional radius $R/R_{\star}$.]{Rotational kernels of the high- and low-order modes of the best-fitting pulsational model as a function of the fractional radius $R/R_{\star}$.}
\label{fig: LowvHighorder}
\end{figure}

We then use this model providing the optimal fit to both the g~modes in the deep stellar interior and the observed p~modes to calculate the rotational frequency implied by the splitting of the $\ell=1$ mode. Under the assumption of slow rotation, a non-radial mode is split due to the Coriolis force into $2\ell+1$ multiplet components with frequencies $f_{\text{nlm}}$ given by the following equation (see e.g. \citealt{Aerts2010}):
\begin{equation}
\hfill f_{\text{nlm}} = f_{\text{nl}}+m(1-C_{\text{nl}})f_{\text{rot, puls}}.\hfill
\label{eq: rotsplit}
\end{equation}
In this equation, $f_{\text{nl}}$ is the unperturbed frequency of the zonal mode, $f_{\text{rot, puls}}$ is the rotational frequency about the pulsational axis and $C_{\text{nl}}$ is the Ledoux constant \citep{Ledoux1951}. We computed it for the best model and obtained a value of $C_{\text{n, }\ell=1}=0.02$. Figure \ref{fig: LowvHighorder} is a display of the rotational kernels of the highest radial order g mode of the best-fitting $\ell=1$ period-spacing pattern, and those of the fundamental radial and dipole g$_1$ mode. It can be inferred from the figure that the high-order g modes have the highest probing power in the near-core region, and the radial and g$_1$ modes have the highest probing power in the stellar envelope.

\begin{figure}[t]
\includegraphics[width=\hsize]{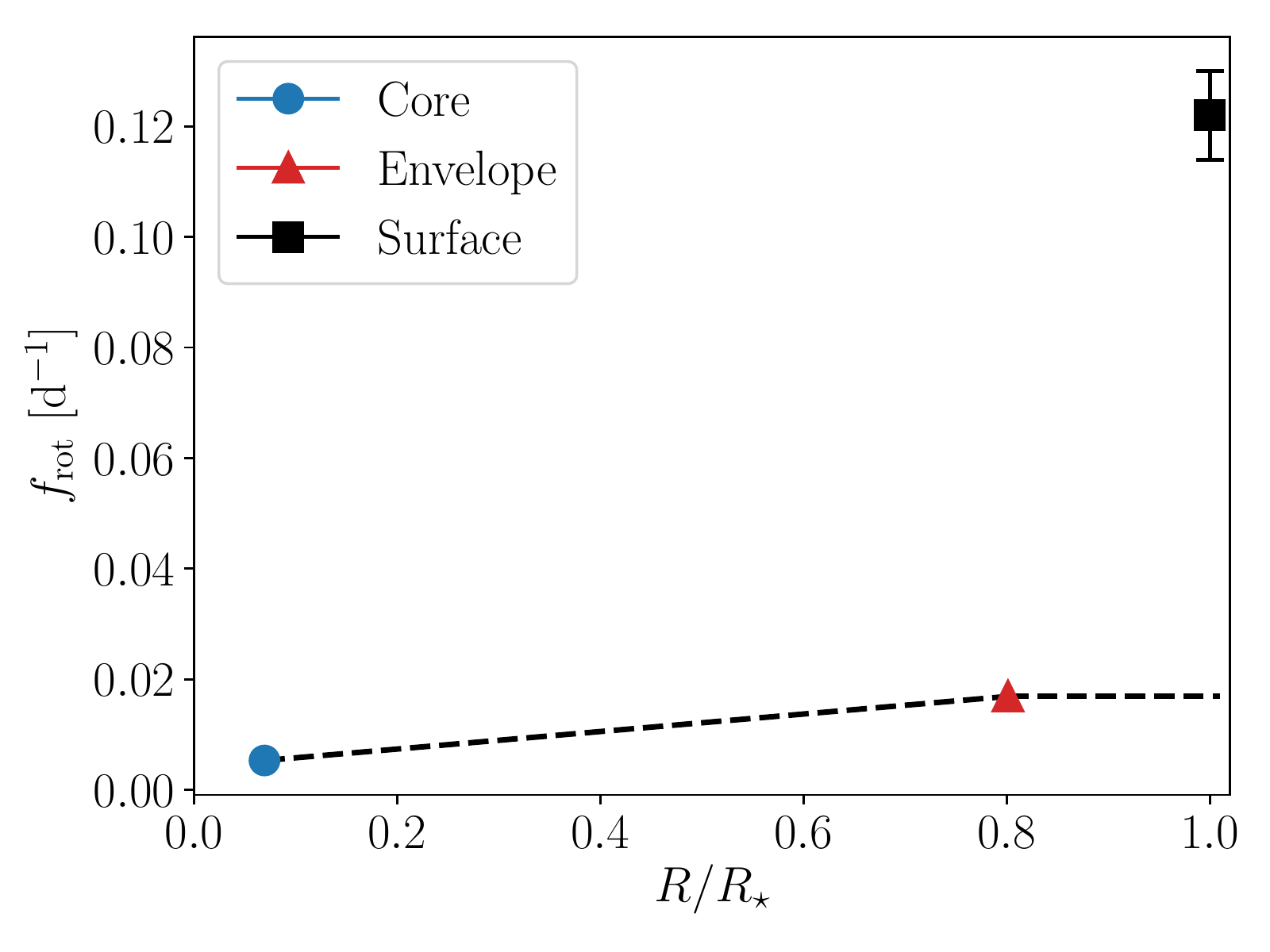}
\caption[Core, envelope and surface rotational frequencies of the primary component of KIC9850387.]{Core (taken from \citealt{Li2020a}, based on the slope of the g-mode period-spacing pattern), envelope (from the rotational splitting of the $\text{g}_{1}$ mode) and surface (from spectroscopic line broadening ignoring pulsational velocity broadening) rotational frequencies} ($f_{\text{rot}}$) as a function of the fractional radius ($R/R_{\star}$) of the primary component of KIC9850387. The core and envelope rotational frequencies are positioned at the maxima of the rotational kernels of the dipole $n_{\text{g}}=39$ and $n_{\text{g}}=1$ modes represented in Figure \ref{fig: LowvHighorder}. The errors on the core and envelope rotational frequencies are smaller than the symbol size. The dashed black line represents the rotational profile from the core to the surface that is compatible with macroturbulent line broadening due pulsations.
\label{fig: Rot_profile}
\end{figure}

\begin{figure}[t]
\includegraphics[width=\hsize]{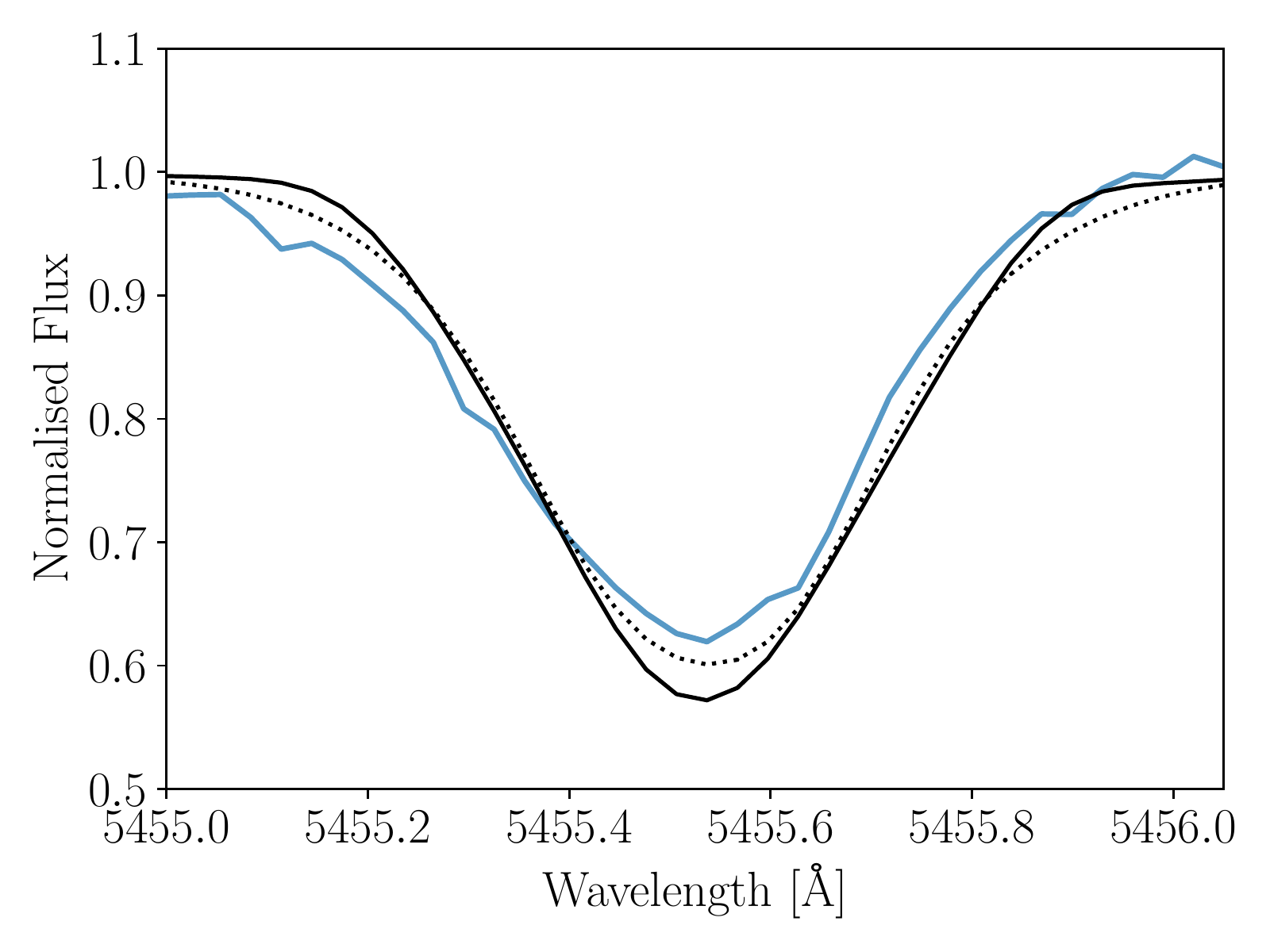}
\caption[Synthetic spectral fits to a Fe I line of the disentangled primary component spectrum with different compositions of rotational broadening.]{Synthetic spectral fits to a Fe I line of the disentangled primary component spectrum with different compositions of rotational broadening. The solid and dotted black lines represent synthetic spectra with identical atmospheric parameter inputs but with ${v\sin{i}=13\text{ km\,s}^{-1}}$ and $v_{\text{macro}}=0\text{ km\,s}^{-1}$, and ${v\sin{i}=2\text{ km\,s}^{-1}}$ and $v_{\text{macro}}=15\text{ km\,s}^{-1}$ respectively.}
\label{fig: Rot_comparison}
\end{figure}

Using Eq.\,(\ref{eq: rotsplit}) and ${f_{\text{nl}}=0.5\cdot (13.21+13.25)=13.23}$~d$^{-1}$ (the midpoint of the doublet), we obtained a value of ${f_{\text{rot, puls}}=0.01689\pm0.00004\text{ d}^{-1}}$.
This value is 3.2 times that of ${f_{\text{rot,core}}=0.0053\pm0.0015\text{ d}^{-1}}$ reported by \cite{Li2020a}, and a factor of 7.2 times lower than the surface rotational frequency $f_{\text{rot, surf}}=0.122_{-0.008}^{+0.008}$~d$^{-1}$ reported in Paper\,I. In that paper, the surface rotation frequency was derived from the projected rotational velocity ($v_{\text{rot}}\sin{i_{\text{orb}}}$) determined through 
spectral line modelling, relying on the orbital inclination ($i_{\text{orb}}$) and radius of the primary star ($R_{\text{p}}$) determined through eclipse modelling, assuming aligned pulsational and orbital axes. This would imply a sharp decrease in the rotational frequency from the surface to the envelope, and a more gradual decrease from the envelope to the near-core region (as displayed in Figure \ref{fig: Rot_profile}). However, the surface rotational frequency deduced from $v_{\text{rot}}\sin{i_{\text{orb}}}$ is necessarily an overestimation of the true surface rotational rate as $v_{\text{rot}}\sin{i_{\text{orb}}}$ was used as a proxy for the total velocity broadening required to fit the spectral lines. The contribution from asymmetric line-profile variations due to the oscillation was ignored (as was mentioned in Paper\,I). Assuming that the envelope rotates rigidly (i.e. ${f_{\text{rot, surf}}=f_{\text{rot, puls}}}$), we obtain ${v_{\text{rot}}\sin{i_{\text{orb}}}=1.823\pm0.005\text{ km\,s}^{-1}}$. Keeping all other atmospheric parameters identical, it was found that the synthetic spectrum generated using a ${v_{\text{rot}}\sin{i_{\text{orb}}}=2\text{ km\,s}^{-1}}$ and including a 
pulsational velocity broadening in the form of macroturbulence following \cite{Aerts2009,Aerts2014}, requires $v_{\text{macro}}=15\text{ km\,s}^{-1}$. Such a profile provides a qualitatively similar fit to the observed profile than the best-fitting synthetic spectrum with ${v_{\text{rot}}\sin{i_{\text{orb}}}=13\text{ km\,s}^{-1}}$ and $v_{\text{macro}}=0\text{ km\,s}^{-1}$ found in Paper\,I, as shown in Figure \ref{fig: Rot_comparison}. Therefore, we conclude that the asteroseismic and spectroscopic data are consistent with a rigidly rotating envelope rotating $\sim 3$ times faster than the deep interior near the convective core.
The near-core region and envelope of the primary star rotate sub-synchronously with respect to the binary orbit by factors
$\sim 69$ and $\sim 22$, respectively.

In the above, we have assumed aligned pulsational and orbital axes. Allowing for misalignment (i.e. ${f_{\text{rot, puls}}=f_{\text{rot, surf}}\sin{(90^{\circ}-i_{\text{offset}})}}$, where $i_{\text{offset}}$ is the angle between the pulsational and orbital axes), we obtain $i_{\text{offset}}=82\pm5^{\circ}$. This would imply that the axes are near orthogonal to each other. While this could be an example of a so-called tidally-trapped or single-sided pulsation \citep{Handler2020,Kurtz2020}, the lack of correlation between these high-frequency modes and the orbital frequency makes this unlikely. Moreover, faster envelope than core rotation has been detected in several other pulsating close binaries (e.g. KIC819776, \citealt{Sowicka2017}, and the inner binary of the triple system HD201433, \citealt{Kallinger2017}). 

One of the potential mechanisms for this effect is the 'inverse' tidal mechanism \citep{Fuller2020}, where tidal interaction with unstable pulsation modes can transfer energy and angular momentum in a manner that forces the star away from synchronicity. The differential rotation of the star, coupled with the asynchrononicity of the surface and envelope rotation with respect to the orbit ($f_{\text{orb}}=0.364$~d$^{-1}$) noted in Paper\,I, seems to reinforce this argument. However, the non-detection of tidally excited or perturbed pulsations (reported in Paper\,I), and the fact that close binaries with synchronous surface rotation but asynchronous core rotation exist (e.g. KIC819776, \citealt{Sowicka2017}) cast doubt on this possibility, though does not rule it out. The second scenario in which a faster envelope than core rotation can develop is by angular momentum transport by internal gravity waves \citep{Rogers2015}, which reinforces our use of mixing profiles calibrated by the theoretical simulations of internal gravity waves in our evolutionary models (see Section \ref{sec: isocloud}).

\section{Discussion and conclusions}
\label{sec: conclusions}

Gravity-mode period-spacing series found in the primary of the eclipsing binary KIC9850387 allowed for high precision estimation of the stellar parameters 
of this intermediate-mass F-type pulsator. We coupled the period spacing patterns of identified dipole and quadrupole modes with an evolutionary and asteroseismic modelling-based analysis by comparing the observationally and theoretically derived parameters of this star. To achieve this goal, we computed a grid of evolutionary models with $0.80\text{ M}_{\odot}\leq M_{\text{ini}}\leq2.00\text{ M}_{\odot}$ with a range of $f_{\text{ov}}$ and $D_{\text{mix}}$ values typical for this type of pulsator \citep{Mombarg2019}, aside from allowing for a broader range of envelope mixing levels to investigate the potential influence of tidal mixing mechanisms in the envelope. We then performed isocloud fitting in a Monte-Carlo framework similar to \cite{Johnston2019a}, and obtained the evolutionary parameters based on the intersection of the dynamical $T_{\text{eff}}$ and log~g constraints of each component with the isocloud parameters corresponding to the 95\% HPD interval of the Monte Carlo age distribution ($\tau_{\text{MC}}=1.3^{+1.5}_{-0.2}$ Gyr).

We exploited the slow-rotating nature of the primary star of KIC9850387 by estimating $\Pi_{0}$ directly from the mean period-spacing $\overline{\Delta P_{\ell}}$ of the $\ell=1$ and $\ell=2$ patterns, and then using it to confront the theoretical $\Pi_{0}$ values extracted from our grids of evolutionary models. Additionally, we modelled the individual $\Delta P_{\ell}$ values of the observed dipole and quadrupole period-spacing patterns by constructing theoretical period-spacing patterns based on non-rotating stellar structural models. We used two different merit functions ($\chi^{2}_{\text{red}}$ and MD) and tested different setups based on the imposition of `pseudo-single-star' spectroscopic and dynamical constraints. It was found that our asteroseismic modelling provided stronger constraints on the interior properties ($M_{\text{cc}}$, $f_{\text{ov}}$ and $D_{\text{mix}}$) of the primary than the evolutionary modelling, demonstrating the probing power of g-modes. We also found an overall agreement between the asteroseismic and evolutionary modelling results.

Our results reinforce the claim of main-sequence binary evolutionary stage made in Paper\,I, contradicting the conclusion of \cite{Zhang2020} that the system comprises two pre-main-sequence components. Our best-fitting models allowed for strong constraints on the parameters describing the interior mixing profile of the star, comprising a low amount of exponentially-decaying core overshooting ($f_{\rm ov}=0.006$) and a high amount of envelope mixing ($D_{\text{mix}}=25$\,cm$^2$\,s$^{-1}$) for this type of pulsator \citep{Mombarg2019}. These findings led to precise constraints on the evolutionary stage of the primary ($X_{\text{c,\ p}}=0.24\pm0.02$) and evolutionary modelling allowed for constraints on the secondary ($X_{\text{c, \ s}}=0.57\pm0.05$), corresponding to an age of $1.2\pm0.1$\,Gyr (see parameters in bold in Table \ref{tab: PS_as_dyn}). We also obtained constraints on the mass of the convective core, $M_{\text{cc}}=0.16\pm0.01$, which is within range of expectation values reported for single F-type g-mode pulsators by \cite{Mombarg2019}.

We exploited the slow-rotating nature of the primary star of KIC9850387 by estimating $\Pi_{0}$ directly from the mean period-spacing $\overline{\Delta P_{\ell}}$ of the $\ell=1$ and $\ell=2$ patterns, and then using it to confront the theoretical $\Pi_{0}$ values extracted from our grids of evolutionary models. Additionally, we modelled the individual period spacing values of the consecutive zonal dipole and quadrupole modes via pulsation computations based on non-rotating equilibrium models.  We used two different merit functions ($\chi^{2}_{\text{red}}$ and MD) and tested different setups based on the imposition of `pseudo-single-star' spectroscopic and dynamical constraints. It was found that our asteroseismic modelling provided stronger constraints on the interior properties ($M_{\text{cc}}$, $f_{\text{ov}}$ and $D_{\text{mix}}$) of the primary than the evolutionary modelling, demonstrating the probing power of g modes. We also found an overall agreement between the asteroseismic and evolutionary modelling results.

We found little difference in the modelling results regardless of whether a $\chi^{2}_{\text{red}}$ or the MD merit function was used and regardless of the grid setup or $\textbf{Y}$ configurations for our $\Pi_{0}$-based modelling. However, the differences are significant for our $\Delta P_{\ell}$-based modelling, particularly when the full grid of models is used in the fit. The application `pseudo-single-star' spectroscopic reduces these discrepancies, and the application of dynamical constraints eliminates them altogether. Due to the degenerate nature of the estimated stellar parameters, the best-fitting period spacing values and errors in the $\chi^{2}_{\text{red}}$ framework are almost identical regardless of the grid subset used in the fit, while those in the MD framework vary significantly as expected from the construction of this merit function.

After identifying the overall best model combining the asteroseismic information of the dipole and quadrupole modes with the dynamical constraints, we investigated the few high-frequency modes that were identified in Paper\,I by calculating the theoretical frequencies of the $\ell=0$ to $\ell=4$ modes in the 12.8~d$^{-1}$ to 13.4~d$^{-1}$ frequency range for our best model. The fundamental radial mode and the ${\text{g}_1}$ mode were found within this frequency range, and frequency offsets of 0.05~d$^{-1}$ and 0.03~d$^{-1}$ between these modes and the nearest observed mode were obtained. We posited that these frequency shifts were due to the surface effect (e.g. \citealt{Ball2017b}), following the explanation of similar behaviour in the binary pulsator KIC10080943  by \cite{Schmid2016}.

We investigated the hypothesis that the observed high-amplitude frequency peaks at 13.21~d$^{-1}$ and 13.25~d$^{-1}$ near the theoretical dipole ${\text{g}_1}$ mode are a part of a rotationally-split prograde-retrograde doublet with a missing zonal mode, as was found in several p- and g-mode hybrid pulsators (e.g. \citealt{Kurtz2014}). This corresponds to an envelope rotational  of ${0.01689\pm0.00004\text{ d}^{-1}}$ that is thrice as high as the core rotational frequency ${0.0053\pm0.0015\text{ d}^{-1}}$ \citep{Li2020a}. Within the limitations of the data, the surface rotation is compatible with the envelope rotation. Similar behaviour has been observed for other close binaries. For single pulsators of the same mass, the faster envelope than core rotation has been explained in terms of angular momentum transport by internal gravity waves triggered by the convective core \citep{Rogers2015}. Such a mechanism may also be active within the primary star of this binary. However, we reinforce that these conclusions are based on the envelope rotational frequency derived from the splitting of a single frequency doublet with posited identification as a dipole mode, and as such is subject to uncertainty.

Overall, we find that asteroseismic theory and observations are only barely compliant, reinforcing the need for homogeneous analyses of samples of pulsating eclipsing binaries that aim at calibrating interior mixing profiles. Such studies would allow for the investigation of the sources of discrepancy between the various parameters, and address weaknesses in the descriptions of angular momentum transport and interior mixing mechanisms.

\begin{acknowledgements}

The research leading to these results has received funding from the Fonds Wetenschappelijk Onderzoek - Vlaanderen (FWO) under the grant agreements G0H5416N (ERC Opvangproject) and G0A2917N  (BlackGEM), and from the European Research Council (ERC) under the European Union’s Horizon 2020 research and innovation programme (grant agreement no. 670519: MAMSIE) and from 
the KU\,Leuven Research Council (grant C16/18/005: PARADISE). The computational resources and services used in this work were provided by the VSC (Flemish Supercomputer Center), funded by the Research Foundation -- Flanders (FWO) and the Flemish Government -- department EWI. The authors are much indebted to the \textsc{mesa} and \textsc{gyre} development teams for making their user-friendly software publicly available. The authors would also like to thank the Leuven MAMSIE team for useful discussions. Last but not least, we would like to acknowledge R. H. D. Townsend for his comments on an earlier version of the manuscript as part of the PhD jury of SS, which led to a more thorough investigation and subsequent refinement of the internal differential rotation results for this star. 

\end{acknowledgements}

\bibliographystyle{aa}
\bibliography{KIC9850387}

\clearpage
\begin{appendix}
\onecolumn
\section{Correlation plots of the MD values in the period-spacing modelling}
\label{sec: correlations}

\begin{figure*}[ht!]
\includegraphics[width=\hsize]{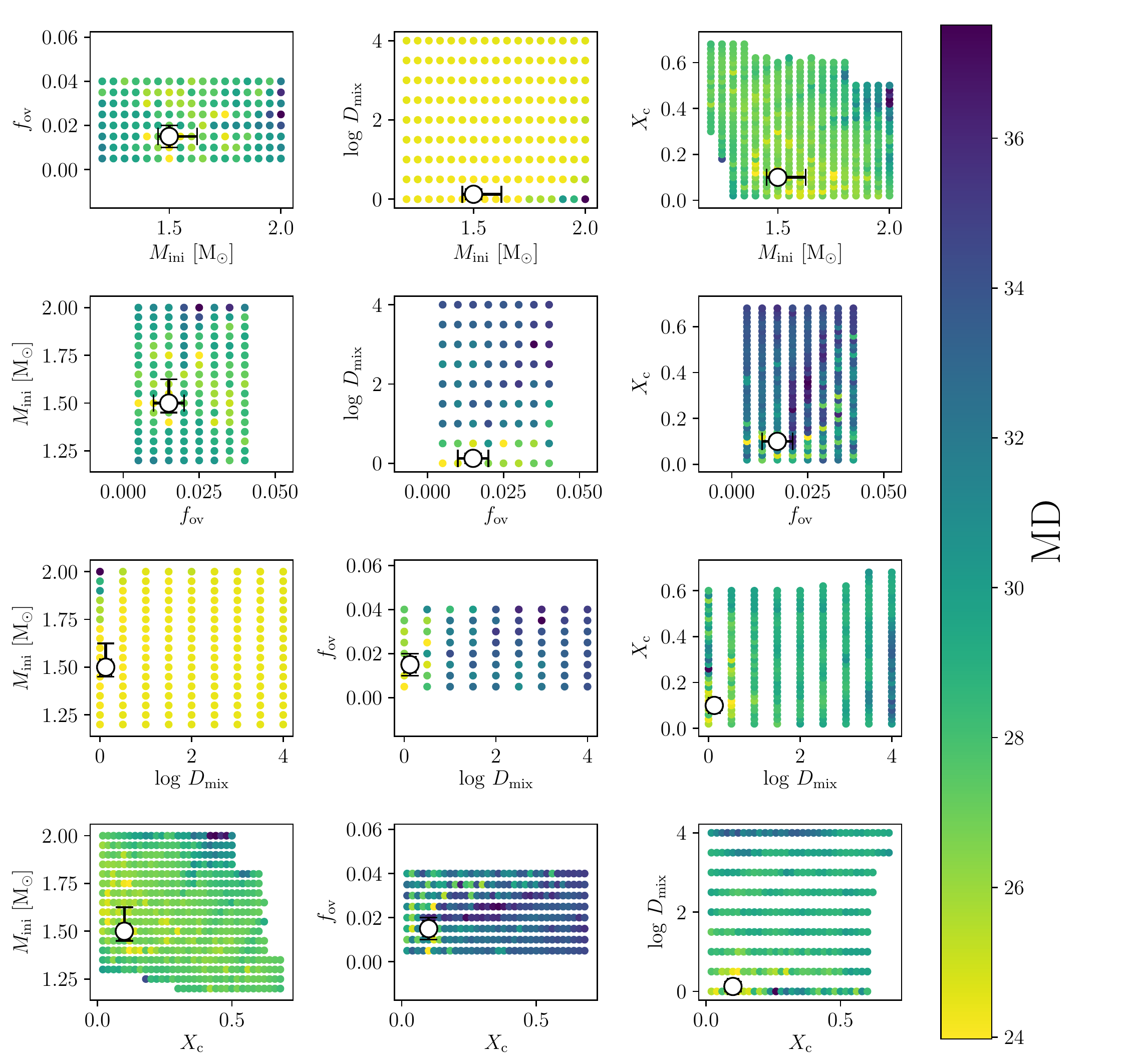}
\caption[Correlations between model parameters for our asteroseismic modelling with $\textbf{Y}=(\Delta P_{\ell=1, \ i})$.]{Correlations between model parameters for our asteroseismic modelling with $\textbf{Y}=(\Delta P_{\ell=1, \ i})$. The points in each subplot are colour-coded according to the MD values of the whole grid. The white circle and error bars represent the maximum-likelihood estimates and half of the 95\% HPD of the Monte Carlo parameter distributions.}
\label{fig: Correlations_l1m0}
\end{figure*}

\newpage
\begin{figure*}[ht!]
\includegraphics[width=\hsize]{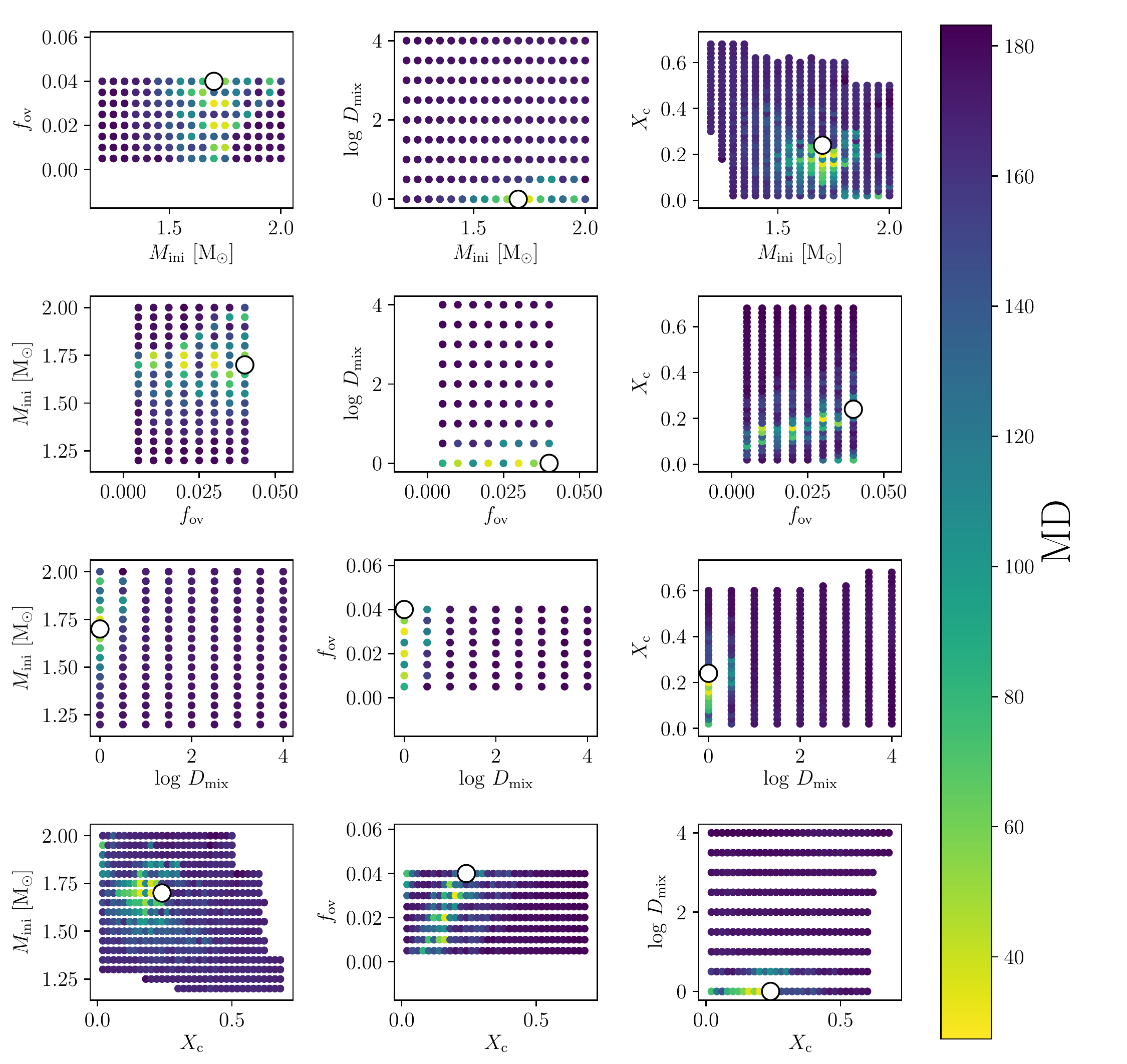}
\caption[Correlations between model parameters when modelling the for our asteroseismic modelling with $\textbf{Y}=(\Delta P_{\ell=2, \ j})$.]{Correlations between model parameters for our asteroseismic modelling with $\textbf{Y}=(\Delta P_{\ell=2, \ j})$. The points in each subplot are colour-coded according to the MD values of the whole grid. The white circle and error bars represent the maximum-likelihood estimates and half of the 95\% HPD of the Monte Carlo parameter distributions.}
\label{fig: Correlations_l2m0}
\end{figure*}

\newpage
\begin{figure*}[ht!]
\includegraphics[width=\hsize]{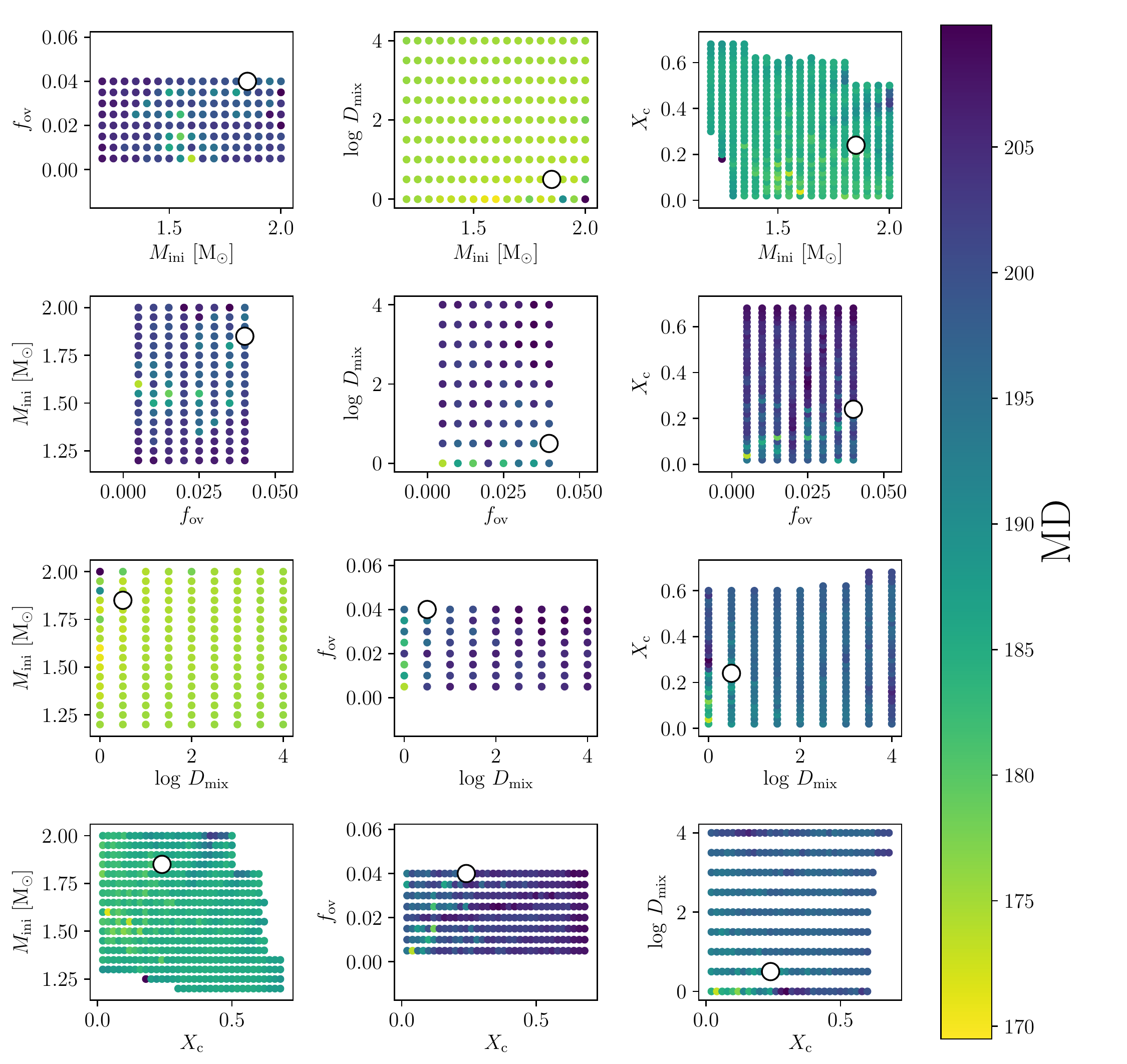}
\caption[Correlations between model parameters for our modelling with $\textbf{Y}=(\Delta P_{\ell=1, \ i}, \Delta P_{\ell=2, \ j})$.]{Correlations between model parameters for our modelling with $\textbf{Y}=(\Delta P_{\ell=1, \ i}, \Delta P_{\ell=2, \ j})$. The points in each subplot are colour-coded according to the MD values of the whole grid. The white circle and error bars represent the maximum-likelihood estimates and half of the 95\% HPD of the Monte-Carlo parameter distributions.}
\label{fig: Correlations_l1m0_l2m0}
\end{figure*}

\newpage
\section{Covariance matrices used in the MD calculations in the period-spacing modelling}
\label{sec: cov_matrices}

\begin{figure*}[!h]
\includegraphics[width=\hsize]{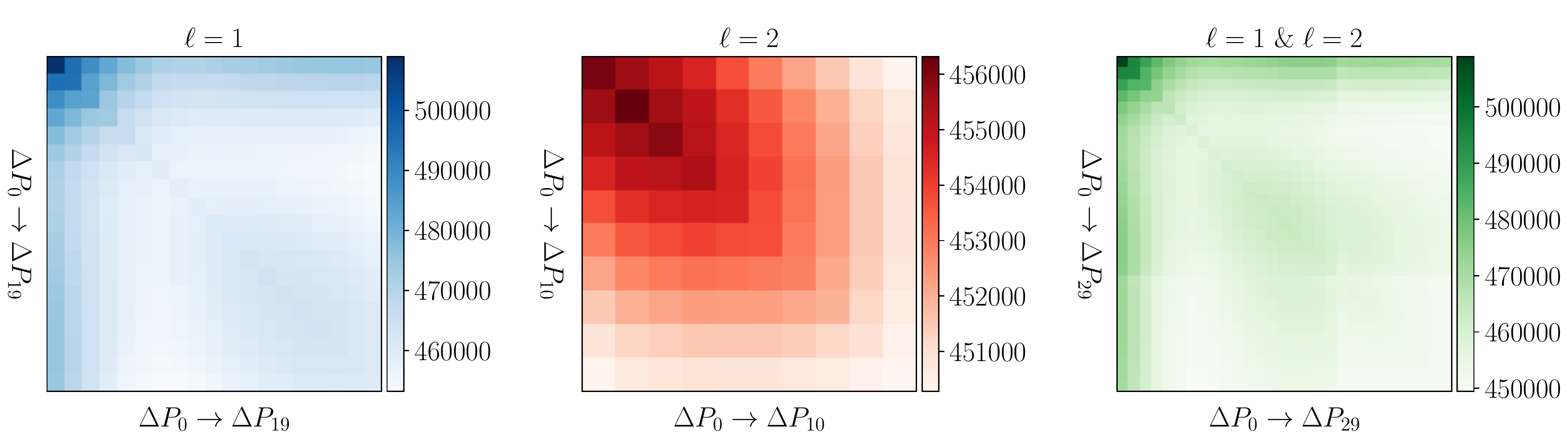}
\caption[Covariance matrices of the individual and combined mode-fitting setups used in the MD calculations in the period-spacing modelling.]{Covariance matrices ($\textbf{V}+\boldsymbol{\Lambda}$, see Eq. \ref{eq: MD}) of the individual and combined mode-fitting setups, representing the 19 $\Delta P_{\ell=1}$ values, the 10 $\Delta P_{\ell=2}$ values of the $\ell=2$ mode, and the 29 total $\Delta P_{\ell=1}$ and $\Delta P_{\ell=2}$ values.}
\label{fig: Cov_matrix}
\end{figure*}

\section{Displacement and rotational-kernel plots for the $\ell=1$ and $\ell=2$ modes of KIC9850387}
\label{sec: disp_kernels}

\begin{figure*}[!h]
\includegraphics[width=\hsize]{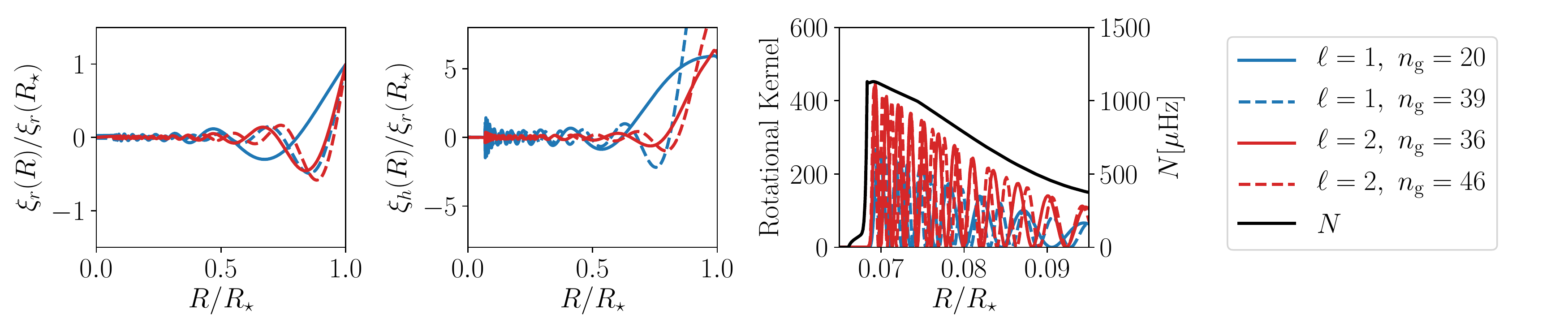}
\caption[Radial ($\xi_{r}(R)$) and horizontal ($\xi_{h}(R)$) components of the Lagrangian displacement vectors, and the rotational kernels.]{\textit{Left} and \textit{middle} panels: Radial ($\xi_{r}(R)$) and horizontal ($\xi_{h}(R)$) components of the Lagrangian displacement vectors. \textit{Right} panel: A plot of the rotational kernels and the Brunt-V{\"a}is{\"a}l{\"a} frequency ($N$) in the near-core region in which the g modes have the highest probing power. These displacement vectors and kernels are of the lowest and highest radial order modes of the theoretical period-spacing pattern of our overall best model ($M=1.65$~M$_{\odot}$, $X_{\text{c}}=0.24$, $f_{\rm ov}=0.005$ and $\log D_{\rm mix}=1.5$). A full derivation of these quantities is provided in \cite{Unno1989} and \cite{Aerts2010}.}
\label{fig: Xi_K}
\end{figure*}

\newpage
\section{Propagation diagrams of the $\ell=1$ and $\ell=2$ modes of KIC9850387}
\label{sec: prop_diag}

\begin{figure}[!h]
\includegraphics[width=\hsize]{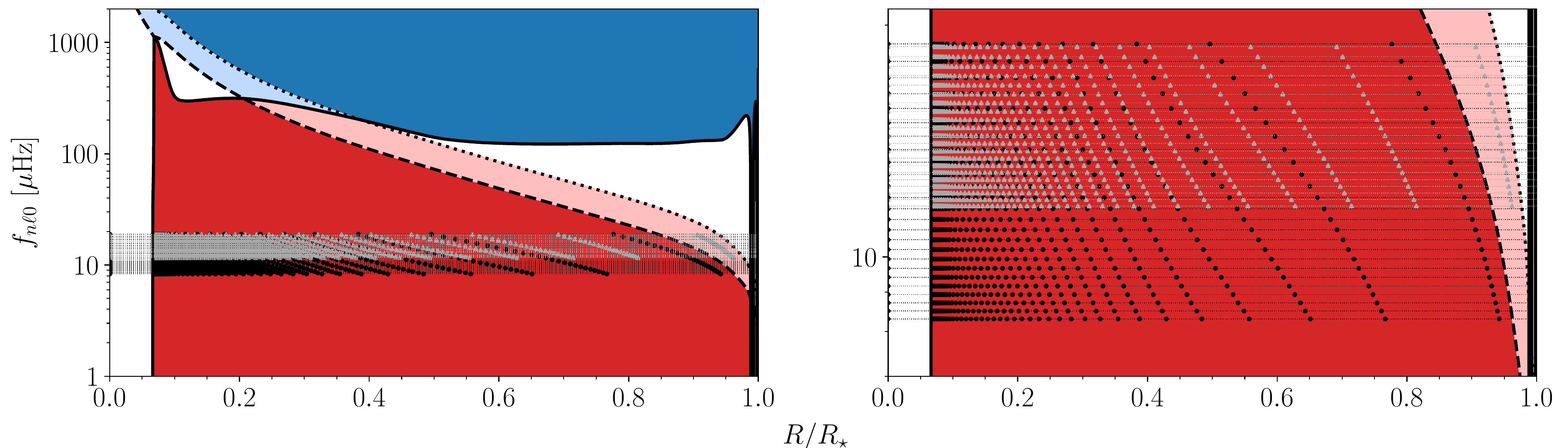}
\caption[Propagation diagrams showing the mode cavities, frequencies and nodes.]{Propagation diagrams showing the mode cavities, frequencies and nodes of the theoretical period-spacing pattern of our overall best model ($M=1.65$~M$_{\odot}$, $X_{\text{c}}=0.24$, $f_{\rm ov}=0.005$ and $\log D_{\rm mix}=1.5$). The thick solid, dashed and dotted curves represent the Brunt-V{\"a}is{\"a}l{\"a} frequency ($N$), and the $\ell=1$ and $\ell=2$ Lamb frequencies ($S_{\ell=1}$ and $S_{\ell=2}$) respectively. The red region represents the $\ell=1$ g mode cavity that gets extended by the pink region for the $\ell=2$ g-modes. The blue region represents the $\ell=2$ p-mode cavity that gets extended by the light blue region for the $\ell=1$ p modes. The horizontal black and grey dotted lines represent the 20 $\ell=1$ and 11 $\ell=2$ theoretical frequencies, with the nodes of radial Lagrangian displacement vector for each frequency represented by black circles and grey triangles respectively. The right panel is a magnification of the frequency region in which the theoretical modes propagate.}
\label{fig: Propagation}
\end{figure}







\end{appendix}

\end{document}